\documentclass[journal]{IEEEtran}
\IEEEoverridecommandlockouts
\usepackage{cite}
\usepackage{amsmath,amssymb,amsfonts}
\usepackage{algorithm}
\usepackage{algpseudocode}
\usepackage{graphicx}
\usepackage{textcomp}
\usepackage{xcolor}
\usepackage{setspace}
\usepackage{gensymb}
\usepackage{diagbox}
\usepackage{adjustbox}
\usepackage{microtype}
\usepackage{multicol,multirow}
\usepackage{stfloats}
\usepackage{mathtools}
\usepackage{subfig}
\usepackage{bbm}
\usepackage{lipsum}

\newenvironment{proof}{{\it Proof:}}{\hfill $\IEEEQED$\par}
\usepackage{caption}
\makeatletter
\newcommand*{\algrule}[1][\algorithmicindent]{\makebox[#1][l]{\hspace*{.5em}\thealgruleextra\vrule height \thealgruleheight depth \thealgruledepth}}%
\newcommand*{\thealgruleextra}{}
\newcommand*{\thealgruleheight}{.75\baselineskip}
\newcommand*{\thealgruledepth}{.25\baselineskip}

\newcount\ALG@printindent@tempcnta
\def\ALG@printindent{%
	\ifnum \theALG@nested>0
	\ifx\ALG@text\ALG@x@notext
	\else
	\unskip
	\addvspace{-1pt}
	\ALG@printindent@tempcnta=1
	\loop
	\algrule[\csname ALG@ind@\the\ALG@printindent@tempcnta\endcsname]%
	\advance \ALG@printindent@tempcnta 1
	\ifnum \ALG@printindent@tempcnta<\numexpr\theALG@nested+1\relax
	\repeat
	\fi
	\fi
}%
\usepackage{etoolbox}
\patchcmd{\ALG@doentity}{\noindent\hskip\ALG@tlm}{\ALG@printindent}{}{\errmessage{failed to patch}}
\makeatother

\newbox\statebox
\newcommand{\myState}[1]{%
	\setbox\statebox=\vbox{#1}%
	\edef\thealgruleheight{\dimexpr \the\ht\statebox+1pt\relax}%
	\edef\thealgruledepth{\dimexpr \the\dp\statebox+1pt\relax}%
	\ifdim\thealgruleheight<.75\baselineskip
	\def\thealgruleheight{\dimexpr .75\baselineskip+1pt\relax}%
	\fi
	\ifdim\thealgruledepth<.25\baselineskip
	\def\thealgruledepth{\dimexpr .25\baselineskip+1pt\relax}%
	\fi
	\State #1%
	\def\thealgruleheight{\dimexpr .75\baselineskip+1pt\relax}%
	\def\thealgruledepth{\dimexpr .25\baselineskip+1pt\relax}%
}

\usepackage[para]{threeparttable}
\AtEndEnvironment{threeparttable}{\vskip10pt}{}{}
\makeatletter
\patchcmd{\TPT@doparanotes}
{\TPT@hsize}
{\TPT@hsize\footnotesize}
{}
{}
\makeatother

\def\BibTeX{\mathbf{B\kern-.05em{\sc i\kern-.025em b}\kern-.08em
		T\kern-.1667em\lower.7ex\hbox{E}\kern-.125emX}}

\begin{document}
	\title{On Minimization/Maximization of the Generalized Multi-Order Complex Quadratic Form With Constant-Modulus Constraints}
	\author{\IEEEauthorblockN{Chunxuan Shi, Yongzhe Li, and Ran Tao}\thanks{The authors are with the School of Information and Electronics, Beijing Institute of Technology, Beijing 100081, China (e-mail: cxshi@bit.edu.cn; yongzhe.li@bit.edu.cn/lyz@ieee.org; rantao@bit.edu.cn).}}
	\maketitle
	\begin{abstract}
		In this paper, we study the generalized problem that minimizes or maximizes a multi-order complex quadratic form with constant-modulus constraints on all elements of its optimization variable. Such a mathematical problem is commonly encountered in various applications of signal processing. We term it as the constant-modulus multi-order complex quadratic programming (CMCQP) in this paper. In general, the CMCQP is non-convex and difficult to solve. Its objective function typically relates to metrics such as signal-to-noise ratio, Cramér-Rao bound, integrated sidelobe level, etc., and constraints normally correspond to requirements on similarity to desired aspects, peak-to-average-power ratio, or constant-modulus property in practical scenarios. In order to find efficient solutions to the CMCQP, we first reformulate it into an unconstrained optimization problem with respect to phase values of the studied variable only. Then, we devise a steepest descent/ascent method with fast determinations on its optimal step sizes. Specifically, we convert the step-size searching problem into a polynomial form that leads to closed-form solutions of high accuracy, wherein the third-order Taylor expansion of the search function is conducted. Our major contributions also lie in investigating the effect of the order and specific form of matrices embedded in the CMCQP, for which two representative cases are identified. Examples of related applications associated with the two cases are also provided for completeness. The proposed methods are summarized into algorithms, whose convergence speeds are verified to be fast by comprehensive simulations and comparisons to existing methods. The accuracy of our proposed fast step-size determination is also evaluated.
	\end{abstract}
	
	\section{Introduction}
	
	\IEEEPARstart{T}{he} optimization on maximizing/minimizing the magnitude of a multi-order complex quadratic form with constant-modulus constraints is of particular significance in signal processing \cite{LevanonRadarSignal,gini2012waveform,StoicaProbingDesing,BluntWFDiversity16,local,robust,TSEBook05,HeWFBook12}. 
	We term such a problem as the \emph{Constant-Modulus Multi-Order Complex Quadratic Programming} (CMCQP). 
	In general, the CMCQP can appear in various applications of radar \cite{LevanonRadarSignal,gini2012waveform,StoicaProbingDesing,BluntWFDiversity16,local,robust}, communications \cite{TSEBook05}, sensing \cite{HeWFBook12}, etc., wherein the research issues such as waveform/code/sequence generation \cite{WicksWF10,HeWFBook12, HeMIMO09,SongWFMaMi16,LiWF18,Shi2022ICASSP}, beamforming/beampattern/beamspace design \cite{bp,bfg}, joint synthesis on transmission and receive filter \cite{SNR0}, and other performance optimizations \cite{CEP2,OFDM,CRB,CRB2} are typically involved. 
	
	In essence, the CMCQP commonly takes a direct or indirect form as its objective function, which can be interpreted as a certain positive-order power of the magnitude on a complex quadratic form. Such an objective function results from some standard metrics elaborated for optimization, depending on the task of interest. As for constraints of the CMCQP, the restriction on all elements of the optimization variable to have constant envelopes (or equivalently, unit magnitudes) is imposed. Such constraints, in most cases, stand for desirable characteristics with respect to applications, which are capable of connecting with performance aspects such as the crest factor \cite{crest}, peak-to-average-power ratio (PAPR) \cite{PAR0}, similarity to desired characteristics \cite{DeMaioPCode09}, uniform amplitude/power allocation accuracy \cite{allo}, to name a few.
	
	The early prototype of the CMCQP can date back to the work of \cite{CMA0} with digital filter design for signal processing, wherein compensations for multipath interference on constant-envelope signal modulations are conducted via filtering. This has enabled signal processing techniques to usher in an era of the so-called constant modulus algorithm (CMA) \cite{CMA2,CMAopt,PAPRR}, which later has been widely applied to wireless communications, in particular, the field of channel equalization. Therein lies the common approximation from the filtered signal to the desired one of constant-modulus, which used to be solved via an unconstrained optimization with a least-square objective. Such early work, however, differs from the CMCQP because of simply incorporating constant-modulus constraints into the objective. By contrast, the CMCQP serves as a more generalized form, which restricts the problem to enable the constant-modulus property via separated constraints. Its popular applications have been evidenced in communications by researches on beamforming \cite{bfg}, constant-envelope precoding \cite{CEP2}, orthogonal frequency division multiplexing (OFDM) \cite{OFDM}, etc.
	
	In the field of radar signal processing or active sensing, the CMCQP can technically deal with optimizations on different metrics including the maximization of  signal-to-noise ratio (SNR) or signal-to-interference-plus-noise ratio (SINR) \cite{SNR0}, the minimization of Cramér-Rao lower bound (CRLB) \cite{CRB,CRB2}, integrated sidelobe level (ISL) \cite{HeWFBook12}, weighted ISL (WISL) \cite{HeMIMO09,SongWFMaMi16,LiWF18,Shi2022ICASSP}, error/distance/divergence of certain orders \cite{CVX}, and so forth. 
	Specifically, the SNR or SINR metric can be optimized by its direct formulation into a quadratic form, which typically involves a covariance matrix of noise and/or interference signals. 
	Instead, the CRLB optimization for the purpose of parameter estimation can be indirectly transformed as a quadratic form by means of establishing relationships with the Fisher information matrix. 
	Moreover, the optimization on ISL/WISL for obtaining good correlation level of waveform(s) at different time lags can be indirectly converted as a piece-wise sum of squares of quadratic forms (i.e., quartic forms).

	Considering that various signal processing applications including the aforementioned ones can boil down to the CMCQP, it is highly necessary to develop a generalized framework for addressing these applications from a mathematical point of view. Unfortunately, such a desired universal framework is missing in literature. Only some special cases of the CMCQP, for example, those studied in \cite{UQPstoica,UQPCui,UQPCui2}, are currently available. They mainly focus on optimizing a real quadratic form with respect to a unimodular vector, and are thereby not suitable for cases that involve modulo operations on complex-valued forms. The methods developed therein are capable of obtaining locally optimal performance via the monotonically error-bound improving technique (see \cite{UQPstoica}), the cyclic manner (see \cite{UQPCui}), or the alternating direction penalty method (see \cite{UQPCui2}). Their big challenges include either high computational complexity or large number of iterations for convergence. In addition, they are also subject to competitive projections from complex values into constant-modulus approximations. These motivate us to find efficient solutions to the CMCQP.

	As for solving the CMCQP, one typical way is to make use of the existing solver or optimization framework, together with certain possible modifications. For example, the semi-definite relaxation (SDR) \cite{Luo2010,SDR0}, majorization/minorization-minimization/maximization (MM) \cite{MM0,MM00}, or gradient descent or gradient ascent (GD/GA) \cite{bfg,mirror,add,GDstep} technique with projections of values on manifold can be applied. 
	In general, the SDR leads to significantly heavy computational complexity, while the MM technique is subject to elaborations on the corresponding majorants/minorants that are difficult to obtain. 
	Therefore, our motivation is also to devise fast solutions to the CMCQP by means of the GD/GA strategy with accelerations \cite{ste2,accEM} and, meanwhile, to avoid repetitive constant-modulus projections for efficiency purposes. 
	
	In terms of the GD/GA regime that will be used to solve the CMCQP, a major issue to address is the determination of step size. This can severely affect the application of the CMCQP in practice \cite{Numerical}. To avoid slow convergence and/or heavy time consumption for the GD/GA, a proper step-size determination system is necessary. Despite some existing work relevant to the issue of step size \cite{CMAopt,GDstep}, they either deal with limited cases free of constant-modulus constraints for the CMCQP \cite{CMAopt}, or conduct an inaccurate step-size approximation \cite{GDstep}. Hence, our motivations additionally include developing unified GD/GA-based frameworks with accurate and fast determination on step sizes for the CMCQP.
	
	In this paper,\footnote{Some preliminary results on the CMCQP have been presented in \cite{Shi2024EUSIPCO}.} we study the problem of minimizing or maximizing a multi-order complex quadratic form with constant-modulus constraints on all elements of its optimization variable.
	We term this problem as the CMCQP, which is commonly encountered in signal processing. 
	Generally, the CMCQP is difficult to solve due to its non-convexity. Our principal goal is to develop efficient algorithms for finding locally stationary points of the CMCQP. The main contributions of our paper are as follows: i) We formulate the generalized optimization form of the CMCQP. This optimization form involves a multi-order modulo operation on a complex quadratic form with an arbitrary matrix embedded, which differs from existing work such as \cite{UQPstoica,UQPCui,UQPCui2}. The major difference lies in whether the involved quadratic form is complex or not; ii) We identify two representative cases for the CMCQP in terms of the specific structure of the matrix involved in the quadratic form, wherein the raised power is properly chosen for both identified cases; iii) We devise two new approaches to solving the CMCQP through direct phase manipulations and efficient GD/GA frameworks with fast and accurate step-size determinations. Both approaches can avoid repetitive projections of complex values into their constant-modulus approximations which are commonly encountered in conventional methods; and iv) For completeness, we provide representative examples of applications and explain the corresponding process to formulate these applications into the CMCQP.
	
	\emph{Notations}: We use bold uppercase, bold lowercase, and italic letters to denote matrices, column vectors, and scalars, respec-tively. Notations $(\cdot)^*$, $(\cdot)^{ \mathrm{ T }}$, $(\cdot)^{ \mathrm{ H }}$, $\odot$, $\otimes$, $\circ$, $|\cdot|$, $\|\cdot\|_{\infty}$, $\|\cdot\|_{\mathrm{F}}$, $\nabla$, $\Re\{\cdot\}$, $\Im\{\cdot\}$ and $\mathrm{vec}(\cdot)$ denote the conjugate, transpose, hermitian, Hadamard product, Kronecker product, Khatri-Rao product, modulus, infinity norm, Frobenius norm, gradient, real part, imaginary part, and column-wise vectorization operations, respectively. Moreover, notations $\mathbb{R}$, $\mathbb{C}$, $\mathbb{N}^{+}$, $\left[\cdot\right]_{m,n}$, $(\cdot)(n)$, $\arg(\cdot)$, $\mathcal{O}(\cdot)$, and $\mathbbm{d}\{\cdot\}$ stand for the real field, complex field, positive integer filed, $(m,n)$-th element of a matrix, $n$-th element of a vector, argument of a complex value, order of a polynomial, and operator that picks up the diagonal entries of a matrix to form a vector, respectively. In addition, $\mathbf{1}_N$ is an $N\times 1$ vector composed of all ones, $\bar{\mathbf{1}}_{n}^{N}$ is the standard $N\times 1$ basis vector whose $n$-th element equals 1 with others all being zeros, and $\mathbf{I}_N$ is the $N\times N$ identity matrix.
	
	\section{Problem Formulation}
	
	Consider the following generalized optimization problem which can either be the minimization or maximization in terms of a complex vector $\mathbf{x}\in\mathbb{C}^{N\times 1}$, i.e., 
	\begin{align}
		\textrm{CMCQP}:\left\{\!\!\!\!\!\!\!
		\begin{array}{ll}
			&\underset{ \mathbf{x} }{\mathrm{min \, (or \, max)}} \quad\! \left|\mathbf{x}^{\mathrm{H}}\mathbf{A}\mathbf{x}\right|^q\\
			&\quad\quad\: \mathrm{s.t.} \qquad\quad  |\mathbf{x}(n)|=c, n=1,\ldots,N
		\end{array}\right.\label{4p3}	
	\end{align}
	wherein the objective function allows for an arbitrary matrix $\mathbf{A}\in\mathbb{C}^{N\times N}$ and involves a $q$-th order ($q>0$) power raised to the modulus of a complex quadratic form with respect to the optimization variable $\mathbf{x}$, and the constraints restrict all the elements of $\mathbf{x}$ to have constant magnitudes equal to $c$. For brevity, we term this problem as the CMCQP and use $c=1$ for derivations without loss of generality.

	Note that the CMCQP given by \eqref{4p3} is non-convex due to the constant-modulus constraints therein. This generalized optimization form differs from relevant existing works such as \cite{UQPstoica,UQPCui,UQPCui2} mainly because of investigating a complex quadratic form associated with any structure of the matrix $\mathbf{A}$. Generally, it is difficult or even impossible to be directly resolved via modern off-the-shelf solvers. Moreover, it does not support the application of conventional techniques such as diagonal loading or symmetrization to convert into unimodular quadratic programming in \cite{UQPstoica,UQPCui,UQPCui2}. To tackle these issues, advanced algorithms are to be developed.
	
	Technically, depending on the explicit structure of $\mathbf{A}$, the CMCQP can lead to different forms. From the perspective of physical meaning, we identify the following two representative cases. The first case (namely, CaseA1) deals with an arbitrary structure of the matrix $\mathbf{A}$ in the objective, which in applications, can relate to the concept such as auto-/cross-correlations between waveforms. The second case (namely, CaseA2) specially involves a Hermitian structure of $\mathbf{A}$ that is additionally positive semi-definite (PSD), whose objective can typically stand for signal-to-noise ratio, Cramér-Rao bound, ambiguity function, etc. For both cases, there is no additional restriction on the value of $q$ except that it is positive, i.e., $q>0$. The following task is to devise specific solutions to CaseA1 and CaseA2 for computational brevity, respectively.
	
	\section{Proposed Gradient Descent/Ascent Methods With Fast and Accurate Step-Size Determination}
	\label{SecGDA}
	
	In this section, we first reformulate the CMCQP into an unconstrained optimization problem. Then, we investigate the effect of the order $q$ on the reformulated CMCQP, which instructs us to carefully select the explicit values of $q$ for both identified CaseA1 and CaseA2. After that, we present our devised GD/GA based approaches with fast step-size determinations for each identified case.
	
	\subsection{Reformulation of the CMCQP}
	
	Let us denote 	$\mathbf{x}\triangleq\left[e^{j\theta_1},\ldots,e^{j\theta_N}\right]^{\mathrm{T}}\in\mathbb{C}^{N\times 1}$ and store all the phase values of $\mathbf{x}$ into a vector given by $\boldsymbol{\theta}\triangleq\left[\theta_1,\ldots,\theta_N\right]^{\mathrm{T}}\in\mathbb{R}^{N\times 1}$ with $\theta_n\in [0,2\pi),n\in\{1,\ldots,N\}$. Hence, we can express the variable $\mathbf{x}$ in \eqref{4p3} as
	\begin{align}
		\mathbf{x}=e^{j\boldsymbol{\theta}}=\cos(\boldsymbol{\theta})+j\sin(\boldsymbol{\theta}) \label{relationship}
	\end{align}
	where $j\triangleq\sqrt{-1}$. Using \eqref{relationship}, we can transform the CMCQP \eqref{4p3} into the form given as follows
	\begin{align}
		\underset{ \boldsymbol{\theta} }{\mathrm{min \, (or \, max)}} 
		\;\;\;
		\left| ( \cos(\boldsymbol{\theta})+j\sin(\boldsymbol{\theta}) )^{\mathrm{H}}\mathbf{A} ( \cos(\boldsymbol{\theta})+j\sin(\boldsymbol{\theta}) ) \right|^q 
		\label{eq:newEq}
	\end{align}
	which is an unconstrained optimization problem with respect to the phase vector $\boldsymbol{\theta}$.
	
	Note that the unconstrained optimization problem \eqref{eq:newEq} still involves complex forms with respect to $\boldsymbol{\theta}$, which can not be solved via current off-the-shelf solvers. To tackle this issue, we come up with the idea of finding the steepest searching direction for \eqref{eq:newEq}. Hence, GD/GA-based methods with fast and accurate determination on step sizes are to be devised.
	
	For the reformulated CMCQP in \eqref{eq:newEq}, it is straightforward that the global optima is independent of the order $q$. Theoretically, we can choose any positive $q$ to address CaseA1 and/or CaseA2 of the CMCQP. However, a poor choice of $q$ may significantly increase the difficulty of developing approaches to solve \eqref{eq:newEq}. Therefore, proper values of $q$ which can facilitate algorithm developments are necessitated for both CaseA1 and CaseA2. With these considerations, in CaseA1, we choose $q=2$ to cooperate with the modulo operation. In CaseA2, we choose $q=1$ since the quadratic form therein is already positively real-valued due to the Hermitian PSD structure of matrix $\mathbf{A}$.
	
	\subsection{Solution to the CMCQP for CaseA1}
	\label{SecGDASubA}
	
	Substituting the above choice $q=2$ to \eqref{eq:newEq}, the optimization problem for CaseA1 can be rewritten as
	\begin{align}
		\label{4pp}
		&\underset{ \boldsymbol{\theta} }{\mathrm{min \, (or \, max)}} \;\;\; \left| ( \cos(\boldsymbol{\theta})+j\sin(\boldsymbol{\theta}) )^{\mathrm{H}}\mathbf{A} ( \cos(\boldsymbol{\theta})+j\sin(\boldsymbol{\theta}) ) \right|^2
	\end{align}
	which is still an unconstrained optimization problem with respect to $\boldsymbol{\theta}$. In order to solve \eqref{4pp}, we choose to develop a GD/GA-based method with fast determination on step sizes.
	
    For brevity, we denote the objective of \eqref{4pp} by $\mathrm{Obj}_{\mathrm{I}}$ hereafter. Its gradient with respect to $\boldsymbol{\theta}$ can be expressed as
		\begin{align}
			\nabla\mathrm{Obj}_{\mathrm{I}} =
			2\Re\{f(\boldsymbol{\theta})\}
			\nabla\Re\{f(\boldsymbol{\theta})\}
			+2\Im\{f(\boldsymbol{\theta})\}\nabla\Im\{f(\boldsymbol{\theta})\}
			\label{4gori}
		\end{align}
		where $f(\boldsymbol{\theta})\triangleq( \cos(\boldsymbol{\theta})+j\sin(\boldsymbol{\theta}) )^{\mathrm{H}}\mathbf{A} ( \cos(\boldsymbol{\theta})+j\sin(\boldsymbol{\theta}) )$, and the fact $\nabla\mathrm{Obj}_{\mathrm{I}}  = \nabla\big(\Re\{f(\boldsymbol{\theta})\}\big)^2+\nabla\big(\Im\{f(\boldsymbol{\theta})\}\big)^2 $ has been used in the derivations. Using \eqref{relationship}, the expression \eqref{4gori} can also be written into a form with respect to $\mathbf{x}$ given by
		\begin{align}
			\label{4gx3}
			\nabla\mathrm{Obj}_{\mathrm{I}}
			&=2\Im\big\{\big(\mathbf{x}^{\mathrm{H}}\mathbf{A}^{\mathrm{H}}\mathbf{x}\mathbf{A}\mathbf{x}+\mathbf{x}^{\mathrm{H}}\mathbf{A}\mathbf{x}\mathbf{A}^{\mathrm{H}}\mathbf{x}\big)\odot \mathbf{x}^*\big\}
		\end{align}
		whose detailed derivations can be found in Appendix~\ref{Proofadd}.
	
	Till now, we can make use of the steepest GD/GA-based strategy \cite{CVX} to solve the reformulated optimization problem \eqref{4pp} for the CMCQP. The remaining task is to find a proper step size $\rho_{\mathrm{I}}$ ($\rho_{\mathrm{I}}\geq0$) for fast and accurately updating $\boldsymbol{\theta}$ at each iteration, which can be expressed as
	\begin{align}
		\boldsymbol{\theta}^{(k+1)}&=\boldsymbol{\theta}^{(k)}+ \tau\rho_{\mathrm{I}}  \nabla^{(k)}\mathrm{Obj}_{\mathrm{I}}\label{up1}
	\end{align}
	where we mark $\boldsymbol{\theta}$ and $\nabla$ with superscript ($k$) (or ($k+1$)) for the $k$-th (or ($k+1$)-th) iteration, and we use $\tau=-1$ and $1$ to stand for the GD and GA cases, respectively. Note that for the GD/GA based update given by \eqref{up1}, 
	it is typically hard to obtain an optimal step with simultaneous good accuracy and low time consumption at each iteration via conventional line-search strategies \cite{CVX,nonl}. To overcome this drawback, we devise a fast and efficient way to determine the step size $\rho_{\mathrm{I}}$ with good accuracy in the following.

	At a certain iteration indexed by $k$, inserting the update process \eqref{up1} into $\mathrm{Obj}_{\mathrm{I}}$ defined below \eqref{4pp}, the optimal step size for the $k$-th iteration can be found by optimization problem given as follows
	\begin{flalign}
		\underset{ \rho_{\mathrm{I}}>0 }{\mathrm{min\,(or\,max)}} \,\, &\big| \big( \cos\big(\boldsymbol{\theta}^{(k)}+\tau\rho_{\mathrm{I}} \nabla^{(k)}\mathrm{Obj}_{\mathrm{I}}\big)+j\sin\big(\boldsymbol{\theta}^{(k)}+\tau\nonumber\\
		&\times \rho_{\mathrm{I}}\nabla^{(k)}\mathrm{Obj}_{\mathrm{I}}\big) \big)^{\mathrm{H}}\mathbf{A} \big( \cos\big(\boldsymbol{\theta}^{(k)}+\tau \nabla^{(k)}\mathrm{Obj}_{\mathrm{I}}\nonumber\\
		&\times\rho_{\mathrm{I}}\big)+j\sin\big(\boldsymbol{\theta}^{(k)}+\tau\rho_{\mathrm{I}} \nabla^{(k)}\mathrm{Obj}_{\mathrm{I}}\big) \big) \big|^2.
		\label{Opt4}
	\end{flalign}
	Note that \eqref{Opt4} is complicated and difficult to tackle because it involves trigonometric functions with the step size $\rho_{\mathrm{I}}$ embedded. Hence, we seek to find an approximate closed-form solution to \eqref{Opt4} with good accuracy. Toward this end, we apply a reasonable Taylor expansion (i.e., the $3$rd-order expansion) to the objective of \eqref{Opt4}, denoted hereafter as $\widetilde{\mathrm{Obj}}^{(k)}_{\mathrm{I}}(\rho_{\mathrm{I}})$ for the $k$-th iteration. Then we have
		\begin{align}
			\label{4objt}
			\widetilde{\mathrm{Obj}}^{(k)}_{\mathrm{I}}(\rho_{\mathrm{I}})
			=\tau\lambda^{(k)}_{\mathrm{I}} \rho_{\mathrm{I}}^3+\mu^{(k)}_{\mathrm{I}} \rho_{\mathrm{I}}^2&+\tau\upsilon^{(k)}_{\mathrm{I}} \rho_{\mathrm{I}}+\mathrm{Obj}^{(k)}_{\mathrm{I}}+\mathcal{O}\big(\rho_{\mathrm{I}}^4\big)
		\end{align}
		where $\mathrm{Obj}^{(k)}_{\mathrm{I}}$ is the value of the objective of \eqref{4pp} obtained at the $k$-th iteration, $\lambda^{(k)}_{\mathrm{I}}$, $\mu^{(k)}_{\mathrm{I}}$, and $\upsilon^{(k)}_{\mathrm{I}}$ are given at the bottom of this page. The detailed derivations to obtain \eqref{4objt} are shown in Appendix~\ref{ProofA}.
	\begin{figure*}[b]
		\hrulefill
		\begin{align}
			\lambda^{(k)}_{\mathrm{I}}=\;&\Im\Big\{\Big(\big(\mathbf{x}^{(k)}\big)^{\mathrm{H}}\mathbf{A}^{\mathrm{H}}\mathbf{x}^{(k)}\mathbf{A}\big(\nabla^{(k)}\mathrm{Obj}_{\mathrm{I}} \odot\mathbf{x}^{(k)}\big)\odot\big(\mathbf{x}^{(k)}\big)^*+\big(\mathbf{x}^{(k)}\big)^{\mathrm{H}}\mathbf{A}\mathbf{x}^{(k)}\mathbf{A}^{\mathrm{H}}\big(\nabla^{(k)}\mathrm{Obj}_{\mathrm{I}}\odot\mathbf{x}^{(k)}\big)\odot\big(\mathbf{x}^{(k)}\big)^*\Big)^\mathrm{T}\big|\nabla^{(k)}\mathrm{Obj}_{\mathrm{I}}\big|^2\nonumber\\
			&
			-\tfrac{1}{3}\Big(\big(\mathbf{x}^{(k)}\big)^{\mathrm{H}}\mathbf{A}^{\mathrm{H}}\mathbf{x}^{(k)}\mathbf{A}\mathbf{x}^{(k)} 
			\odot\big(\mathbf{x}^{(k)}\big)^*+\big(\mathbf{x}^{(k)}\big)^{\mathrm{H}}\mathbf{A}\mathbf{x}^{(k)}\mathbf{A}^{\mathrm{H}}\mathbf{x}^{(k)}\odot\big(\mathbf{x}^{(k)}\big)^*\Big)^\mathrm{T}\big|\nabla^{(k)}\mathrm{Obj}_{\mathrm{I}}\big|^3-\Big(\big(\mathbf{A}^\mathrm{H}\mathbf{x}^{(k)}
			\odot\big(\mathbf{x}^{(k)}\big)^*\nonumber\\
			&
			+\mathbf{A}\big(\mathbf{x}^{(k)}\big)^*\odot\mathbf{x}^{(k)}\big)^{\mathrm{T}}\big|\nabla^{(k)}\mathrm{Obj}_{\mathrm{I}}\big|^2-2\big(\mathbf{A}\big(\nabla^{(k)}\mathrm{Obj}_{\mathrm{I}}\odot\big(\mathbf{x}^{(k)}\big)^*\big)\odot\mathbf{x}^{(k)}\big)^{\mathrm{T}}\nabla^{(k)}\mathrm{Obj}_{\mathrm{I}}\Big)\nonumber\\
			&\times\big(\mathbf{A}\mathbf{x}^{(k)}
			\odot\big(\mathbf{x}^{(k)}\big)^*-\mathbf{A}^{\mathrm{H}}\big(\mathbf{x}^{(k)}\big)^*\odot\mathbf{x}^{(k)}\big)^{\mathrm{T}}\nabla^{(k)}\mathrm{Obj}_{\mathrm{I}}\Big\}\label{mu3}\\
			\mu^{(k)}_{\mathrm{I}}=\;&\Re\Big\{\big(\nabla^{(k)}\mathrm{Obj}_{\mathrm{I}}\big)^{\mathrm{T}}\Big(\big(\mathbf{x}^{(k)}\big)^{\mathrm{H}}\mathbf{A}^{\mathrm{H}}\mathbf{x}^{(k)}\mathbf{A}\big(\nabla^{(k)}\mathrm{Obj}_{\mathrm{I}}
			\odot\mathbf{x}^{(k)}\big)\odot\big(\mathbf{x}^{(k)}\big)^*+\big(\mathbf{x}^{(k)}\big)^{\mathrm{H}}\mathbf{A}\mathbf{x}^{(k)}\mathbf{A}^{\mathrm{H}}\big(\nabla^{(k)}\mathrm{Obj}_{\mathrm{I}}\odot\mathbf{x}^{(k)}\big)\odot\big(\mathbf{x}^{(k)}\big)^*\Big)\nonumber\\
			&
			-\big(\big|\nabla^{(k)}\mathrm{Obj}_{\mathrm{I}}\big|^2\big)^{\mathrm{T}}\Big(\big(\mathbf{x}^{(k)}\big)^{\mathrm{H}}\mathbf{A}^{\mathrm{H}}\mathbf{x}^{(k)}\mathbf{A}\mathbf{x}^{(k)}\odot\big(\mathbf{x}^{(k)}\big)^*+\big(\mathbf{x}^{(k)}\big)^{\mathrm{H}}\mathbf{A}\mathbf{x}^{(k)}\mathbf{A}^{\mathrm{H}}\mathbf{x}^{(k)}\odot\big(\mathbf{x}^{(k)}\big)^*\Big)\Big\}\nonumber\\
			&
			+\big|\big(\mathbf{A}\mathbf{x}^{(k)}\odot\big(\mathbf{x}^{(k)}\big)^*-\mathbf{A}^{\mathrm{H}}\big(\mathbf{x}^{(k)}\big)^*\odot\mathbf{x}^{(k)}\big)^\mathrm{T}\nabla^{(k)}\mathrm{Obj}_{\mathrm{I}}\big|^2\label{mu2}\\
			\upsilon^{(k)}_{\mathrm{I}}=\;&\big(\nabla^{(k)}\mathrm{Obj}_{\mathrm{I}}\big)^{\mathrm{T}}\nabla^{(k)}\mathrm{Obj}_{\mathrm{I}}\label{mu1}
		\end{align}
	\end{figure*}
	
	Ignoring the constant and high-order terms which are immaterial to the optimization,\footnote{The effect of high-order terms in the expansions can be ignored if the step size satisfies $\rho_{\mathrm{I}}\ll 1/\|\nabla^{(k)}\mathrm{Obj}_{\mathrm{I}}\|$. In this case, the high-order terms become extremely small, which can be easily guaranteed by the proper choice of $\rho_{\mathrm{I}}$.} the step-size search problem \eqref{Opt4} can be rewritten as
	\begin{align}
		\underset{ \rho_{\mathrm{I}}>0 }{\mathrm{min\,(or\,max)}} \quad \tau\lambda^{(k)}_{\mathrm{I}} \rho_{\mathrm{I}}^3+\mu^{(k)}_{\mathrm{I}} \rho_{\mathrm{I}}^2+\tau\upsilon^{(k)}_{\mathrm{I}} \rho_{\mathrm{I}}
		\label{Opt4e}
	\end{align}
	whose solution is the positive root to the derivative of the objective, i.e.,
	\begin{align}
		\rho_{\mathrm{I}}^{(k)}&= \frac{-\mu^{(k)}_{\mathrm{I}}-\tau\sqrt{\big(\mu^{(k)}_{\mathrm{I}}\big)^2-3\lambda^{(k)}_{\mathrm{I}}\upsilon^{(k)}_{\mathrm{I}}}}{3\tau\lambda^{(k)}_{\mathrm{I}}}.
		\label{t14}
	\end{align}
	In the case of no positive root to the derivative of the objective of \eqref{Opt4e}, i.e., the right-hand side of \eqref{t14} becomes negative or complex, a modified small step size of user choice can be enforced. For example, we can choose the following 
	\begin{align}
		\rho_{\mathrm{I}}^{(k)}=\frac{2}{\beta_{\mathrm{max}}\big(\nabla^2\mathrm{Obj}_{\mathrm{I}}(\boldsymbol{\theta}^{(k)})\big)}\label{stepcorrect}
	\end{align}
	to guarantee the convergence for iterations (see analysis in Sec. III-D), where $\beta_{\mathrm{max}}(\cdot)$ is the largest eigenvalue of a matrix.

	The proposed algorithm based on GD/GA with optimal step-size determination for CaseA1 is concluded in Algorithm 1. In practice, Algorithm 1 can be technically fast by removing the repetitive computations that occur in $\nabla^{(k)}\mathrm{Obj}_{\mathrm{I}}$, $\lambda^{(k)}_{\mathrm{I}}$, and $\mu^{(k)}_{\mathrm{I}}$, wherein the common components for fast calculations can be expressed as
	\begin{align}
		\mathbf{s}^{(k)}_{\mathrm{I}}&\triangleq\big(\mathbf{x}^{(k)}\big)^{\mathrm{H}}\mathbf{A}^{\mathrm{H}}\mathbf{x}^{(k)}\mathbf{A}\mathbf{x}^{(k)}\odot\big(\mathbf{x}^{(k)}\big)^*\nonumber\\
		& \qquad\qquad\qquad+\big(\mathbf{x}^{(k)}\big)^{\mathrm{H}}\mathbf{A}\mathbf{x}^{(k)}\mathbf{A}^{\mathrm{H}}\mathbf{x}^{(k)}\odot\big(\mathbf{x}^{(k)}\big)^*
	\end{align}
    \begin{align}
		\mathbf{t}^{(k)}_{\mathrm{I}}&\triangleq\big(\mathbf{x}^{(k)}\big)^{\mathrm{H}}\mathbf{A}^{\mathrm{H}}\mathbf{x}^{(k)}\mathbf{A}\big(\nabla^{(k)}\mathrm{Obj}_{\mathrm{I}}\odot\mathbf{x}^{(k)}\big)\odot\big(\mathbf{x}^{(k)}\big)^*\nonumber\\
		&+\big(\mathbf{x}^{(k)}\big)^{\mathrm{H}}\mathbf{A}\mathbf{x}^{(k)}\mathbf{A}^{\mathrm{H}}\big(\nabla^{(k)}\mathrm{Obj}_{\mathrm{I}}\odot\mathbf{x}^{(k)}\big)\odot\big(\mathbf{x}^{(k)}\big)^*.\!
	\end{align}	
	
	Based on the above implementations, we can analyze the per iteration computational complexity of Algorithm 1. Specifically, the calculations of $\mathbf{s}^{(k)}_{\mathrm{I}}$ and $\mathbf{t}^{(k)}_{\mathrm{I}}$ require $4N^2+4N$ and $4N^2+6N$ operations, respectively. Moreover, the calculations of $\lambda^{(k)}_{\mathrm{I}}$, $\mu^{(k)}_{\mathrm{I}}$, and $\upsilon^{(k)}_{\mathrm{I}}$ by means of $\mathbf{s}^{(k)}_{\mathrm{I}}$ and $\mathbf{t}^{(k)}_{\mathrm{I}}$ are respectively $5N^2+15N$, $2N^2+6N$, and $N$ operations. In addition, the update from $\mathbf{x}^{(k)}$ to $\mathbf{x}^{(k+1)}$ requires $N$ operations. Therefore, the overall calculations of each iteration in Algorithm 1 cost $15N^2+33N$ operations, meaning that the per iteration computational complexity of Algorithm 1 is $\mathcal{O}(N^2)$.

	\begin{figure}[t]%
		\vspace{-4pt}
		\begin{algorithm}[H]
			\caption{The proposed algorithm design to address the CMCQP for CaseA1 \big(Min/Max-CMCQP$^{\text{I}}$\big).}
			\label{alg2}
			\begin{algorithmic}[1]
				\vspace{-3pt}
				\myState {Initialization: $\boldsymbol{\theta}^{(0)}$, $\mathbf{x}^{(0)}=e^{j\boldsymbol{\theta}^{(0)}}$, $k\leftarrow 0$}
				\Repeat {}
				\myState
				{Calculate $\nabla^{(k)}\mathrm{Obj}_{\mathrm{I}}$ via \eqref{4gx3} by enforcing $\mathbf{x}=\mathbf{x}^{(k)}$}
				\myState{Calculate $\lambda^{(k)}_{\mathrm{I}}$, $\mu^{(k)}_{\mathrm{I}}$, and $\upsilon^{(k)}_{\mathrm{I}}$ via \eqref{mu3}-\eqref{mu1}}
				\myState{Calculate $\rho_{\mathrm{I}}^{(k)}$ via \eqref{t14}}
				\myState{Calculate $\boldsymbol{\theta}^{(k+1)}$ via \eqref{up1}}
				\myState{$\mathbf{x}^{(k+1)}=e^{j\boldsymbol{\theta}^{(k+1)}}$}
				\myState{$ k \leftarrow k+1$}
				\Until convergence
			\end{algorithmic}
		\end{algorithm}
		\vspace{-18pt}
	\end{figure}
	
	\subsection{Solution to the CMCQP for CaseA2}
	
	Since the matrix $\mathbf{A}$ for CaseA2 has a Hermitian PSD structure and the order of interest therein is $q=1$, we can remove the modulo operation in \eqref{eq:newEq}. Thus, the CMCQP associated with CaseA2 can be rewritten as
	\begin{align}
		\label{2pp}
		&\underset{ \boldsymbol{\theta} }{\mathrm{min \, (or \, max)}} \;\;\; ( \cos(\boldsymbol{\theta})+j\sin(\boldsymbol{\theta}) )^{\mathrm{H}}\mathbf{A} ( \cos(\boldsymbol{\theta})+j\sin(\boldsymbol{\theta}) )
	\end{align}
	which is still an unconstrained optimization problem with a compound quadratic objective.
	
	In order to solve \eqref{2pp}, we also devise a GD/GA based design with fast determinations on step sizes for iterations using the same idea as shown in the previous subsection. For brevity, we denote the objective of \eqref{2pp} as $\mathrm{Obj}_{\mathrm{II}}$, whose gradient can be obtained by
	\begin{align}
		\nabla\mathrm{Obj}_{\mathrm{II}}&=2\big(\Im\{\mathbf{A}\}\cos(\boldsymbol{\theta})+\Re\{\mathbf{A}\}\sin(\boldsymbol{\theta})\big)\odot\cos{(\boldsymbol{\theta})}\nonumber\\
		&+2\big(\Im\{\mathbf{A}\}\sin(\boldsymbol{\theta})-\Re\{\mathbf{A}\}\cos(\boldsymbol{\theta})\big)\odot\sin{(\boldsymbol{\theta})}.\label{2g2}
	\end{align}
	
	To further simplify the calculation and avoid separating the real and imaginary parts in \eqref{2g2}, we seek to recover $\nabla\mathrm{Obj}_{\mathrm{II}}$ into a form expressed in terms of $\mathbf{x}$. For this purpose, we apply \eqref{relationship} to rewrite $\nabla\mathrm{Obj}_{\mathrm{II}}$ in \eqref{2g2} as follows
	\begin{align}
		\nabla\mathrm{Obj}_{\mathrm{II}}=2\Im\{\mathbf{A}\mathbf{x}\odot\mathbf{x^*}\}.\label{g2}
	\end{align}
	
	Till now, we can solve the unconstrained optimization problem \eqref{2pp} on the basis of \eqref{g2}. Similar to CaseA1, we also make use of the steepest GD/GA-based strategy and seek to find a proper step size $\rho_{\mathrm{II}}$ ($\rho_{\mathrm{II}}\geq0$) for fast and accurately updating $\boldsymbol{\theta}$ at the $k$-th iteration, i.e.,
	\begin{align}
		\boldsymbol{\theta}^{(k+1)}&=\boldsymbol{\theta}^{(k)}+ \tau\rho_{\mathrm{II}} \nabla^{(k)}\mathrm{Obj}_{\mathrm{II}}\label{up11}
	\end{align}
	where $\tau$ has been defined in \eqref{up1}.
	
	At the $k$-th iteration, inserting the update process \eqref{up11} into $\mathrm{Obj}_{\mathrm{II}}$ which has been defined before, the steepest-step-size search problem for CaseA2 can be cast as
	\begin{flalign}
		\underset{ \rho_{\mathrm{II}}>0 }{\mathrm{min\,(or\,max)}} \, & \big( \cos\big(\boldsymbol{\theta}^{(k)}+\tau\rho_{\mathrm{II}}\nabla^{(k)}\mathrm{Obj}_{\mathrm{II}}\big)+j\sin\big(\boldsymbol{\theta}^{(k)}+\tau\nonumber\\
		&\times\rho_{\mathrm{II}}\nabla^{(k)}\mathrm{Obj}_{\mathrm{II}}\big) \big)^{\mathrm{H}}\mathbf{A} \big( \cos\big(\boldsymbol{\theta}^{(k)}+\nabla^{(k)}\mathrm{Obj}_{\mathrm{II}}\nonumber\\
		&\times\tau\rho_{\mathrm{II}}\big)+j\sin\big(\boldsymbol{\theta}^{(k)}+\tau\rho_{\mathrm{II}}\nabla^{(k)}\mathrm{Obj}_{\mathrm{II}}\big) \big).\!\!\!
		\label{Opt2}
	\end{flalign}
	
	In order to solve the optimization problem \eqref{Opt2} efficiently, our strategy is still to conduct the third-order Taylor expansion on its objective function (denoted hereafter as $\widetilde{\mathrm{Obj}}^{(k)}_{\mathrm{II}}(\rho_{\mathrm{II}})$). The expansion result with the proof given in Appendix~\ref{ProofB} can be expressed as follows
	\begin{align}
		\label{2objt}
		\widetilde{\mathrm{Obj}}^{(k)}_{\mathrm{II}}(\rho_{\mathrm{II}})
		=\;&\tau\lambda^{(k)}_{\mathrm{II}} \rho_{\mathrm{II}}^3+\mu^{(k)}_{\mathrm{II}} \rho_{\mathrm{II}}^2+\tau\upsilon^{(k)}_{\mathrm{II}} \rho_{\mathrm{II}}+\mathrm{Obj}^{(k)}_{\mathrm{II}}\nonumber\\
		&+\mathcal{O}\big(\rho_{\mathrm{II}}^4\big)
	\end{align}
	where $\mathrm{Obj}^{(k)}_{\mathrm{II}}$ is the value of $\mathrm{Obj}_{\mathrm{II}} $ obtained at the $k$-th iteration, and $ \lambda^{(k)}_{\mathrm{II}} $, $ \mu^{(k)}_{\mathrm{II}} $, and $ \upsilon^{(k)}_{\mathrm{II}} $ are respectively given by
	\begin{flalign}
		\lambda^{(k)}_{\mathrm{II}}=\;&\Im\Big\{\big(\mathbf{A}\big(\nabla^{(k)}\mathrm{Obj}_{\mathrm{II}}\odot\mathbf{x}^{(k)}\big)\odot\big(\mathbf{x}^{(k)}\big)^*\big)^\mathrm{T}\big|\nabla^{(k)}\mathrm{Obj}_{\mathrm{II}}\big|^2\nonumber\\
		& -\tfrac{1}{3}\big(\mathbf{A}\mathbf{x}^{(k)}\odot\big(\mathbf{x}^{(k)}\big)^*\big)^\mathrm{T}\big|\nabla^{(k)}\mathrm{Obj}_{\mathrm{II}}\big|^3\Big\}\label{l23}\\
		\mu^{(k)}_{\mathrm{II}}
		=\;&\Re\Big\{\big(\mathbf{A}\big(\nabla^{(k)}\mathrm{Obj}_{\mathrm{II}}\odot\mathbf{x}^{(k)}\big)\odot\big(\mathbf{x}^{(k)}\big)^*\big)^\mathrm{T}\nabla^{(k)}\mathrm{Obj}_{\mathrm{II}}\nonumber\\
		& -\big(\mathbf{A}\mathbf{x}^{(k)}\odot\big(\mathbf{x}^{(k)}\big)^*\big)^\mathrm{T}\big|\nabla^{(k)}\mathrm{Obj}_{\mathrm{II}}\big|^2\Big\}\label{l22}\\
		\upsilon^{(k)}_{\mathrm{II}}
		=\;&\big(\nabla^{(k)}\mathrm{Obj}_{\mathrm{II}}\big)^{\mathrm{T}}\nabla^{(k)}\mathrm{Obj}_{\mathrm{II}} .
		\label{l21}
	\end{flalign}
	
	Ignoring the constant and high-order terms with respect to $\rho_{\mathrm{II}}$ in \eqref{2objt},\footnote{The same condition as presented in \eqref{4objt} is applied here.} we can rewrite the step-size search problem \eqref{Opt2} into the form as follows
	\begin{align}
		\underset{ \rho_{\mathrm{II}}>0 }{\mathrm{min\,(or\,max)}} \;\;\; \tau\lambda^{(k)}_{\mathrm{II}} \rho_{\mathrm{II}}^3+\mu^{(k)}_{\mathrm{II}} \rho_{\mathrm{II}}^2+\tau\upsilon^{(k)}_{\mathrm{II}} \rho_{\mathrm{II}}
		\label{Opt2e}
	\end{align}
	whose solution is the positive root to the derivative of its objective function, i.e.,
	\begin{align}
		\rho_{\mathrm{II}}^{(k)}&= \frac{-\mu^{(k)}_{\mathrm{II}}-\tau\sqrt{\big(\mu^{(k)}_{\mathrm{II}}\big)^2-3\lambda^{(k)}_{\mathrm{II}}\upsilon^{(k)}_{\mathrm{II}}}}{3\tau\lambda^{(k)}_{\mathrm{II}}}.
		\label{t12}
	\end{align}
	In the case of no positive root to the derivative of the objective of \eqref{Opt2e}, i.e., the right-hand side of \eqref{t12} becomes negative or complex, a modified small step size of user choice given in \eqref{stepcorrect} can be implemented by replacing $\rho_{\mathrm{I}}^{(k)}$ and $\nabla^{(k)}\mathrm{Obj}_{\mathrm{I}}$ with $\rho_{\mathrm{II}}^{(k)}$ and $\nabla^{(k)}\mathrm{Obj}_{\mathrm{II}}$, respectively.

	The proposed algorithm based on GD/GA with optimal step-size determination for CaseA2 is concluded in Algorithm 2, whose implementation can also be accelerated by removing repetitive computations with same forms. To fulfill this goal, we identify the common parts in the calculations of $\nabla^{(k)}\mathrm{Obj}_{\mathrm{II}}$, $\lambda^{(k)}_{\mathrm{II}}$, and $\mu^{(k)}_{\mathrm{II}}$, which are listed as follows
	\begin{align}
		\mathbf{s}^{(k)}_{\mathrm{II}}&\triangleq\mathbf{A}\mathbf{x}^{(k)}\odot\big(\mathbf{x}^{(k)}\big)^*\\
		\mathbf{t}^{(k)}_{\mathrm{II}}&\triangleq\mathbf{A}\big(\nabla^{(k)}\mathrm{Obj}_{\mathrm{II}}\odot\mathbf{x}^{(k)}\big)\odot\big(\mathbf{x}^{(k)}\big)^*.
	\end{align} 
	
	Regarding the per iteration computational complexity of Algorithm 2, the calculations of $\mathbf{s}^{(k)}_{\mathrm{II}}$ and $\mathbf{t}^{(k)}_{\mathrm{II}}$ require $N^2+N$ and $N^2+2N$ operations, respectively. The computations of $\lambda^{(k)}_{\mathrm{II}}$, $\mu^{(k)}_{\mathrm{II}}$, and $\upsilon^{(k)}_{\mathrm{II}}$ respectively cost $5N$, $3N$, and $N$ operations via $\mathbf{s}^{(k)}_{\mathrm{II}}$ and $\mathbf{t}^{(k)}_{\mathrm{II}}$. In addition, the update of $\mathbf{x}$ needs $N$ operations. Consequently, the overall per iteration cost of Algorithm 2 is $2N^2+13N$ operations, whose computational complexity is also $\mathcal{O}(N^2)$.
	
	\begin{figure}[!t]%
		\vspace{-4pt}
		\begin{algorithm}[H]
			\caption{The proposed algorithm design to address the CMCQP for CaseA2 \big(Min/Max-CMCQP$^{\text{II}}$\big).}
			\label{alg1}
			\begin{algorithmic}[1]
				\vspace{-3pt}
				\myState {Initialization: $\boldsymbol{\theta}^{(0)}$, $\mathbf{x}^{(0)}=e^{j\boldsymbol{\theta}^{(0)}}$, $k\leftarrow 0$}
				\Repeat {}
				\myState{Calculate $\nabla^{(k)}\mathrm{Obj}_{\mathrm{II}}$ via \eqref{g2} by replacing $\mathbf{x}$ with $\mathbf{x}^{(k)}$}
				\myState{Calculate $\lambda^{(k)}_{\mathrm{II}}$, $\mu^{(k)}_{\mathrm{II}}$, and $\upsilon^{(k)}_{\mathrm{II}}$ via \eqref{l23}-\eqref{l21}}
				\myState{Calculate $\rho_{\mathrm{II}}^{(k)}$ via \eqref{t12}}
				\myState{Calculate $\boldsymbol{\theta}^{(k+1)}$ via \eqref{up11}}
				\myState{$\mathbf{x}^{(k+1)}=e^{j\boldsymbol{\theta}^{(k+1)}}$}
				\myState{$ k \leftarrow k+1$}
				\Until convergence
			\end{algorithmic}
		\end{algorithm}
		\vspace{-18pt}
	\end{figure}
	
	\subsection{Convergence Analysis}
	
	The local convergence of our proposed algorithms Min/Max-CMCQP$^{\mathrm{I}}$ and Min/Max-CMCQP$^{\mathrm{II}}$ can be guaranteed on the condition that the step-size determinations \eqref{t14} and \eqref{t12} are of good accuracy. Technically, we conduct the following analysis for Algorithms 1 and 2. Here, we omit the subscripts $(\cdot)_{\mathrm{I}}$ and $(\cdot)_{\mathrm{II}}$ to present common results for both CaseA1 and CaseA2.
	
	Let us consider the minimization case for example. According to \eqref{Opt4} and \eqref{Opt2}, our proposed step size is obtained by the approximation $\rho^{(k)}\approx{\mathrm{argmin}_{\rho}}\;\mathrm{Obj}(\boldsymbol{\theta}^{(k)}-\rho \nabla^{(k)} \mathrm{Obj})$. When such an approximation is highly accurate, based on \eqref{up1} and \eqref{up11}, we can obtain $\mathrm{Obj}^{(k+1)}=\mathrm{Obj}(\boldsymbol{\theta}^{(k)}-\rho^{(k)} \nabla^{(k)} \mathrm{Obj})\leq\mathrm{Obj}(\boldsymbol{\theta}^{(k)}-0\cdot \nabla^{(k)} \mathrm{Obj})=\mathrm{Obj}^{(k)}$. Then, the monotone non-increasing property of the objective across iterations can be ensured. Once the non-increasing trend deteriorates at a certain point, a backup scheme of step-size determination can be triggered, which can be given by \eqref{stepcorrect}. Using \eqref{stepcorrect} and the fact
		\begin{align}
			\!{\mathrm{Obj}}(\boldsymbol{\theta})\leq {\mathrm{Obj}}^{(k)}&+\Re\big\{(\boldsymbol{\theta}-\boldsymbol{\theta}^{(k)})^\mathrm{H}\nabla^{(k)}\mathrm{Obj}\big\}\nonumber\\
			&+\tfrac{1}{2}\beta_{\mathrm{max}}\big(\nabla^2\mathrm{Obj}(\boldsymbol{\theta}^{(k)})\big)\big\|\boldsymbol{\theta}-\boldsymbol{\theta}^{(k)}\big\|^2\label{seTy}\!
		\end{align}
		by enforcing $\boldsymbol{\theta}\triangleq\boldsymbol{\theta}^{(k+1)}$, we can obtain $\mathrm{Obj}^{(k+1)}\leq\mathrm{Obj}^{(k)}$.
		
		To find the conditions of convergence, we further make a reasonable assumption that the objective function $\mathrm{Obj}$ is lower bounded by an infimum denoted as $\mathcal{L}(\mathrm{Obj})$. Then, based on the definition of infimum, we know that $\exists K\in\mathbb{N}^{+}, \forall \epsilon>0, \mathcal{L}(\mathrm{Obj})+\epsilon>\mathrm{Obj}^{(K)}$. Using the monotone non-increasing property of $\mathrm{Obj}$, we also have $\forall k>K, \mathrm{Obj}^{(K)}\geq \mathrm{Obj}^{(k)} > \mathcal{L}(\mathrm{Obj})-\epsilon$. Combining the above two results, we can obtain the following fact 
		\begin{align}
			\exists K\in\mathbb{N}^{+}, \forall \epsilon>0, \mathcal{L}(\mathrm{Obj})-\epsilon< \mathrm{Obj}^{(k)} <\mathcal{L}(\mathrm{Obj})\;+\;&\epsilon,\nonumber\\
			\forall k>K &\label{objcon}
		\end{align}
		which is identical to the definition for the convergence of $\mathrm{Obj}$ across iterations.
		
		In order to further prove that the gradient $\nabla\mathrm{Obj}$ can converge to zeros, we first re-express \eqref{objcon} as the limitation form, i.e., 
		\begin{align}
			\underset{k\rightarrow +\infty}{\lim} \big(\mathrm{Obj}(\boldsymbol{\theta}^{(k)}-\rho\nabla^{(k)}\mathrm{Obj})-\mathrm{Obj}(\boldsymbol{\theta}^{(k)})\big)=0,&\nonumber\\
			\forall \rho\leq{2}/{\beta_{\mathrm{max}}\big(\nabla^2\mathrm{Obj}(\boldsymbol{\theta}^{(k)})\big)}&\label{limi}
		\end{align}
		where we have used \eqref{seTy} to determine the range of $\rho$. The limitation \eqref{limi} indicates that the value of $\mathrm{Obj}$ keeps the same on the line connecting $\boldsymbol{\theta}^{(k)}$ and $\boldsymbol{\theta}^{(k)}-\frac{2\nabla^{(k)}\mathrm{Obj}}{\beta_{\mathrm{max}}(\nabla^2\mathrm{Obj}(\boldsymbol{\theta}^{(k)}))}$ for a sufficiently large $k$, which enforces $\nabla^{(k)}\mathrm{Obj}$ to only be zeros if $\mathrm{Obj}$ is first-order continuous with respect to $\boldsymbol{\theta}$. The same routine can be applied to the maximization case.
		
		Finally, we give the conditions for the local convergence of our algorithms as follows: i) the objective $\mathrm{Obj}$ is bounded; ii) the objective $\mathrm{Obj}$ is second-order differentiable and is first-order continuous; and iii) we use the backup step size \eqref{stepcorrect} when the non-increasing trend of objective deteriorates by enforcing \eqref{t14} or \eqref{t12} at a certain iteration.
	
	\section{Applications of the CMCQP}
	\subsection{CaseA1 of the CMCQP for Waveform Design}
	
	The CMCQP is applicable to the field of waveform(s)/code(s) design that requires all the elements to have constant or unit energy. Without loss of generality, we assume that a set of $ M $ unimodular waveforms is to be designed, each of which has a code length $ P $. Here, we denote the $m$-th waveform for transmission by the vector $ \mathbf{y}_{m} \triangleq \left[ e^{ j \psi_{m} (1)}, \ldots, e^{ j \psi_{m} (P)} \right]^{ \mathrm{ T } } \in \mathbb{C}^{ P \times  1 }  $, wherein all the phase values $   \psi_{m} (p), \, p = 1, \ldots, P  $ arbitrarily range between $-\pi$ and $\pi$. We store all the $ M $ waveforms into a matrix denoted by $ \mathbf{Y} \triangleq [\mathbf{y}_1,\ldots,\mathbf{y}_M] \in \mathbb{C}^{ P \times M  }  $.
	
	In the context of multi-waveform design, one classical way is to synthesize waveforms with low auto- and cross-correlations, i.e., to enable low integrated sidelobe level or WISL of waveforms. To this end, a generalized WISL minimization-based design problem can be formulated. Mathematically, it can be expressed as follows (see (8) of \cite{Shi2022ICASSP})
	\begin{align}
		\label{optProbWF}
		\underset{ \mathbf{ Y } } {  \mathrm{ min } }
		&
		\quad
		\sum_{m'=1}^{M}\sum_{m=1}^{M}\sum_{p=-P+1}^{P-1}
		\gamma_{p}^{2}\left|\mathbf{y}_{m'}^{ \mathrm{ H }}\mathbf{J}_p \mathbf{y}_{m}\right|^2\nonumber
		\\
		\mathrm{ s.t. }
		&\quad
		|[\mathbf{Y}]_{m,p}|=1, \; m=1, \ldots, M;\,p=1,\ldots,P
	\end{align}
	where $ \{ \gamma_{p} \}_{p=-P+1}^{P} $ are ISL controlling weights which normally take values $0$ or $1$, and $\{ \mathbf{J}_p \}_{p=-P+1}^{P-1} $ are shift matrices (Positive/negative $ p $ for upper/lower shift, and $ \mathbf{J}_{0} =\mathbf{I}_{P} $).
	
	We let $ \mathbf{y} \triangleq \mathrm{vec} ( \mathbf{Y} ) = [ \mathbf{y}_{1}^{ \mathrm{ T } }, \ldots, \mathbf{y}_{M}^{ \mathrm{ T } } ]^{ \mathrm{ T } } $, and therefore can transform the problem \eqref{optProbWF} into a form that falls within the scope of the CMCQP given by
	\begin{align}
		\label{design}
		\underset{ {\mathbf{ y }} } {  \mathrm{ min } }
		&
		\quad
		\sum_{m'=1}^{M}\sum_{m=1}^{M}\sum_{p=-P+1}^{P-1}\big|\mathbf{ y }^{\mathrm{H}}\mathbf{A}_{m'mp}\mathbf{ y }\big|^2\nonumber
		\\
		\mathrm{ s.t. }
		&\quad
		\left| \mathbf{ y }(n)  \right| = 1, \; n=1, \ldots, MP
	\end{align}
	where we introduce the matrices $ \mathbf{A}_{m'mp} \triangleq \gamma_{p}\big(\bar{\mathbf{1}}^M_{m'}(\bar{\mathbf{1}}_{m}^{M})^{\mathrm{T}}\big)\otimes\mathbf{J}_p \in \mathbb{R}^{ MP \times MP  },  m',m\in\{1,\ldots,M\}; p \in \{ -P+1, \ldots, P-1 \}$ for the reformulation.
	
	Note that \eqref{design} takes the form of a piece-wise sum of the CMCQP for CaseA1. To solve it using the proposed solution in Sec.~\ref{SecGDASubA}, we denote the objective function of \eqref{design} by $ \mathrm{Obj}_{\mathrm{WF}} \triangleq \sum_{m'=1}^{M}\sum_{m=1}^{M} \sum_{p=-P+1}^{P-1} \mathrm{Obj}_{\mathrm{I},m'mp} $ where the $(m',m,p)$-th piece-wise objective $ \mathrm{Obj}_{\mathrm{I},m'mp} $ is defined as $\mathrm{Obj}_{\mathrm{I},m'mp}\triangleq\big|(\cos(\boldsymbol{\theta})+j\sin(\boldsymbol{\theta}))^{\mathrm{H}}\mathbf{A}_{m'mp}(\cos(\boldsymbol{\theta})+j\sin(\boldsymbol{\theta}))\big|^2 $. Applying \eqref{4gx3} to each piece-wise component $\mathrm{Obj}_{\mathrm{I},m'mp}, \, \forall m',m\in\{1,\ldots,M\};\,\forall p \in \{ -P+1, \ldots, P-1 \}$, we can obtain the gradient of $ \mathrm{Obj}_{\mathrm{WF}} $ at the $k$-th iteration as
	\begin{flalign}
		&
		\nabla^{(k)}\mathrm{Obj}_{\mathrm{WF}}
		=
		\sum_{m'=1}^{M}\sum_{m=1}^{M}\sum_{\substack{p=-P+1}}^{P-1} \nabla^{(k)}\mathrm{Obj}_{\mathrm{I},m'mp}\nonumber
		\\
		&
		\ =
		\sum_{m'=1}^{M}\sum_{m=1}^{M}\sum_{\substack{p=-P+1}}^{P-1}
		2\Im\Big\{\Big(\big(\mathbf{ y }^{(k)}\big)^{\mathrm{H}}\mathbf{A}_{m'mp}^{\mathrm{H}}\mathbf{ y }^{(k)}\mathbf{A}_{m'mp}\mathbf{ y }^{(k)}\nonumber\\
		&
		\quad\quad
		+\big(\mathbf{ y }^{(k)}\big)^{\mathrm{H}}\mathbf{A}_{m'mp}\mathbf{ y }^{(k)}\mathbf{A}_{m'mp}^{\mathrm{H}}\mathbf{ y }^{(k)}\Big)\odot \big(\mathbf{ y }^{(k)}\big)^*\Big\}
		\label{gwa1}
	\end{flalign}
	where $\mathbf{ y }^{(k)}$ stands for the update of the overall waveform vector obtained at the $k$-th iteration.
	
	The remaining task is to calculate the polynomial coefficients, which are also composed of each piece-wise component associated with the matrix $ \mathbf{A}_{m'mp}, \, \forall m',m\in\{1,\ldots,M\}; \, \forall p \in \{ -P+1, \ldots, P-1 \} $. 	Using $ \nabla^{(k)}\mathrm{Obj}_{\mathrm{WF}} $ obtained from \eqref{gwa1} and also the matrix $ \mathbf{A}_{m'mp} $, by enforcing $  \nabla^{(k)}\mathrm{Obj}_{\mathrm{I}} = \nabla^{(k)}\mathrm{Obj}_{\mathrm{WF}} $ and $ \mathbf{A} = \mathbf{A}_{m'mp} $ to \eqref{mu3}--\eqref{mu1}, we can arrive to the expressions of the parameters for step-size determination given by \eqref{muwave3}--\eqref{muwave1}, respectively, which are shown at the bottom of the next page.
	
	\addtocounter{equation}{3}
	
	The common components to avoid repetitive calculations are respectively given by
	\begin{align}
		\mathbf{s}_{\mathrm{WF}}^{(k)}\triangleq\;&\sum_{m'=1}^{M}\sum_{m=1}^{M}\sum_{\substack{p=-P+1}}^{P-1}\Big(\big(\mathbf{ y }^{(k)}\big)^{\mathrm{H}}\mathbf{A}_{m'mp}^{\mathrm{H}}\mathbf{ y }^{(k)}\mathbf{A}_{m'mp}\mathbf{ y }^{(k)}\nonumber\\
		&+\big(\mathbf{ y }^{(k)}\big)^{\mathrm{H}}\mathbf{A}_{m'mp}\mathbf{ y }^{(k)}\mathbf{A}_{m'mp}^{\mathrm{H}}\mathbf{ y }^{(k)}\Big)\odot \big(\mathbf{ y }^{(k)}\big)^*\label{s0}
	\end{align}
    
    \begin{align}
    	\nonumber\\[-11mm]
		\mathbf{t}_{\mathrm{WF}}^{(k)}\triangleq\;&\sum_{m'=1}^{M}\sum_{m=1}^{M}\sum_{\substack{p=-P+1}}^{P-1}\Big(\big(\mathbf{ y }^{(k)}\big)^{\mathrm{H}}\mathbf{A}_{m'mp}^{\mathrm{H}}\mathbf{ y }^{(k)}\mathbf{A}_{m'mp}\nonumber\\
		&\; \times\big(\nabla^{(k)}\mathrm{Obj}_{\mathrm{WF}}\odot\mathbf{ y }^{(k)}\big)+\big(\mathbf{ y }^{(k)}\big)^{\mathrm{H}}\mathbf{A}_{m'mp}\mathbf{ y }^{(k)}\nonumber\\
		&\; \times\mathbf{A}_{m'mp}^{\mathrm{H}}\big(\nabla^{(k)}\mathrm{Obj}_{\mathrm{WF}}\odot\mathbf{ y }^{(k)}\big)\Big)\odot \big(\mathbf{ y }^{(k)}\big)^*.\label{s1}
	\end{align}

	\begin{figure}[!t]%
		\vspace{-4pt}
		\begin{algorithm}[H]
			\caption{The proposed algorithm design to address the CMCQP for waveform design \big(Min-CMCQP$^{\text{WF}}$\big).}
			\label{alg3}
			\begin{algorithmic}[1]
				\vspace{-3pt}
				\myState {Initialization: $\boldsymbol{\theta}^{(0)}$, $\mathbf{y}^{(0)}=e^{j\boldsymbol{\theta}^{(0)}}$, $k\leftarrow 0$}
				\Repeat {}
				\myState
				{Calculate $\nabla^{(k)}\mathrm{Obj}_{\mathrm{WF}}$ via \eqref{gwa1}}
				\myState
				{Calculate $\lambda^{(k)}_{\mathrm{WF}}$, $\mu^{(k)}_{\mathrm{WF}}$, and $\upsilon^{(k)}_{\mathrm{WF}}$ via \eqref{muwave3}-\eqref{s1}}			
				\myState{$\rho_{\mathrm{WF}}^{(k)}=\frac{-\mu^{(k)}_{\mathrm{WF}}+\sqrt{\big(\mu^{(k)}_{\mathrm{WF}}\big)^2-3\lambda^{(k)}_{\mathrm{WF}}\upsilon^{(k)}_{\mathrm{WF}}}}{-3\lambda^{(k)}_{\mathrm{WF}}}$}
				\myState{$\boldsymbol{\theta}^{(k+1)}=\boldsymbol{\theta}^{(k)}-\rho^{(k)}_{\mathrm{WF}}\nabla^{(k)}\mathrm{Obj}_{\mathrm{WF}}$}
				\myState{$\mathbf{ y }^{(k+1)}=e^{j\boldsymbol{\theta}^{(k+1)}}$}
				\myState{$ k \leftarrow k+1$}
				\Until convergence
				\myState {Recover $\mathbf{Y}$ from $\mathbf{ y }$}
			\end{algorithmic}
		\end{algorithm}
		\vspace{-18pt}
	\end{figure}

    \begin{figure*}[b]
    	\hrulefill
    	\addtocounter{equation}{-5}
    	\begin{align}
    		\lambda^{(k)}_{\mathrm{WF}}=&\sum_{m'=1}^{M}\sum_{m=1}^{M}\sum_{\substack{p=-P+1}}^{P-1}\Im\Big\{
    		\Big(2\big(\mathbf{A}_{m'mp}\big(\nabla^{(k)}\mathrm{Obj}_{\mathrm{WF}}\odot\big(\mathbf{ y }^{(k)}\big)^*\big)\odot\mathbf{ y }^{(k)}\big)^\mathrm{T}
    		\nabla^{(k)}\mathrm{Obj}_{\mathrm{WF}}-\big(\mathbf{A}_{m'mp}^{\mathrm{T}}\mathbf{ y }^{(k)}\odot\big(\mathbf{ y }^{(k)}\big)^*
    		\nonumber\\
    		&+\mathbf{A}_{m'mp}\big(\mathbf{y}^{(k)}\big)^*\odot\mathbf{y}^{(k)}\big)^\mathrm{T}\big|\nabla^{(k)}\mathrm{Obj}_{\mathrm{WF}}\big|^2\Big)\big(\mathbf{A}_{m'mp}\mathbf{ y }^{(k)}
    		\odot\big(\mathbf{ y }^{(k)}\big)^*-\mathbf{A}_{m'mp}^{\mathrm{T}}\big(\mathbf{ y }^{(k)}\big)^*\odot\mathbf{ y }^{(k)}\big)^\mathrm{T}\nabla^{(k)}\mathrm{Obj}_{\mathrm{WF}}\Big\}\nonumber\\
    		&+\Im\Big\{\big(\mathbf{t}^{(k)}_{\mathrm{WF}}\big)^{\mathrm{T}}\big|\nabla^{(k)}\mathrm{Obj}_{\mathrm{WF}}\big|^2-\tfrac{1}{3}\big(\big|\nabla^{(k)}\mathrm{Obj}_{\mathrm{WF}}\big|^3\big)^\mathrm{T}
    		\mathbf{s}^{(k)}_{\mathrm{WF}}\Big\}\label{muwave3}\\
    		\mu^{(k)}_{\mathrm{WF}}
    		=& \sum_{m'=1}^{M}\sum_{m=1}^{M}\sum_{\substack{p=-P+1}}^{P-1}\big|\big(\mathbf{A}_{m'mp}\mathbf{ y }^{(k)}\odot\big(\mathbf{ y }^{(k)}\big)^*-\mathbf{A}_{m'mp}^{\mathrm{T}}\big(\mathbf{ y }^{(k)}\big)^*\odot\mathbf{ y }^{(k)}\big)^{\mathrm{T}}\nabla^{(k)}\mathrm{Obj}_{\mathrm{WF}}\big|^2\nonumber\\
    		&+\Re\Big\{\big(\mathbf{t}_{\mathrm{WF}}^{(k)}\big)^{\mathrm{T}}\nabla^{(k)}\mathrm{Obj}_{\mathrm{WF}}-\big(\mathbf{s}_{\mathrm{WF}}^{(k)}\big)^{\mathrm{T}}\big|\nabla^{(k)}\mathrm{Obj}_{\mathrm{WF}}\big|^2\Big\}\label{muwave2}\\
    		\upsilon^{(k)}_{\mathrm{WF}}=\;&\big(\nabla^{(k)}\mathrm{Obj}_{\mathrm{WF}}\big)^\mathrm{T}\nabla^{(k)}\mathrm{Obj}_{\mathrm{WF}}\label{muwave1}
    	\end{align}
    \end{figure*}
    \addtocounter{equation}{2}
	
	Till now, we can conclude the application of the CMCQP for waveform design and summarize the corresponding procedures in Algorithm 3.
	
	\subsection{CaseA2 of the CMCQP for Metric Optimizations}
	
	The CMCQP with CaseA2 is also applicable to various fields of signal processing for radar, communications, as well as the recently emerged issue on their joint co-existence. In essence, the CMCQP with CaseA2 is typically suitable for certain metric optimizations, which can include the beampattern/beamspace energy ratio \cite{bp}, SINR \cite{SNR0}, CRLB \cite{CRB2}, PAPR \cite{PAPRR}, ambiguity function (AF) \cite{AF0}, to name a few. These metrics can establish relationships to specific functions of applications such as beampattern/beamspace design, code synthesis, parameter estimation, AF shaping, etc. As for constant-modulus constraints on the optimization variable of the CMCQP with CaseA2, they are normally introduced by the application itself.
	
	Without loss of generality, we show the explicit expression of the CMCQP with CaseA2 in the form as follows
	\begin{align}
		\underset{ \mathbf{y} }{\mathrm{min \, (or \, max)}}
		&\quad
		\mathbf{y}^\mathrm{H}\mathbf{A}\mathbf{y}\nonumber
		\\
		\quad\quad\: \mathrm{s.t.} \qquad
		&\quad
		|\mathbf{y}(n)|=1, \; n=1, \ldots, N
		\label{app2}
	\end{align}
	which can be a direct formulation for any of the aforementioned metric optimizations, or instead, an indirect reformulation after some manipulations such as the elementary transform or change on modeling variables. In terms of applications, the Hermitian PSD matrix $\mathbf{A}$ depends on specific scenarios, toward which some representative examples are shown in the following.
	
	\subsubsection{Optimum Detection Design}
	
	In the context of target detection for radar or radar-communications with spectrally dense environment, the optimum detection design (e.g., the generalized likelihood ratio test) is typically converted to the maximization of SNR or SINR at the receiving end \cite{SNR0}. In this case, the corresponding design problem can be recast as a maximization form of \eqref{app2}. Specifically, the original SNR or SINR maximization problem can be expressed as
	\begin{align}
		\label{designSINR}
		\underset{ \mathbf{w},\mathbf{z} } {  \mathrm{ max } }
		& \quad \frac{|\alpha|^2\cdot|\mathbf{w}^\mathrm{H}(\mathbf{z}\odot\mathbf{d})|
			^2}{\mathbf{w}^\mathrm{H}\mathbf{R}\mathbf{w}}
		\quad
		\nonumber
		\\
		\mathrm{ s.t. }
		&\quad
		\left| \mathbf{ z }(n)  \right| = 1, \; n=1, \ldots, N
	\end{align}
	where $\alpha$ and $\mathbf{d}\in\mathbb{C}^{N\times1}$ are the complex reflection coefficient and Doppler steering vector of the target, respectively, $\mathbf{R}\in\mathbb{C}^{N\times N}$ is the covariance matrix of noise or signal-independent interference plus noise, and $\mathbf{w}\in\mathbb{C}^{N\times1}$ and $\mathbf{z}\in\mathbb{C}^{N\times1}$ are vectors of adaptive filter weights and polyphase codes to be designed, respectively.
	
	For fixed $\mathbf{z}$, the solution to \eqref{designSINR} obeys the minimum variance distortionless response (MVDR) expression which can be given by the following form
	\begin{align}
		\mathbf{w}=\frac{\mathbf{R}^{-1}(\mathbf{z}\odot\mathbf{d})}{(\mathbf{z}\odot\mathbf{d})^\mathrm{H}\mathbf{R}^{-1}(\mathbf{z}\odot\mathbf{d})}.\label{MVDR}
	\end{align}
	Substituting the MVDR solution \eqref{MVDR} into the objective of \eqref{designSINR} and ignoring constant terms which are immaterial to optimization, we can transform \eqref{designSINR} into the form as follows
	\begin{align}
		\label{designSINR2}
		\underset{ \mathbf{z} } {  \mathrm{ max } }
		& \quad (\mathbf{z}\odot\mathbf{d})^\mathrm{H}\mathbf{R}^{-1}(\mathbf{z}\odot\mathbf{d})
		\quad
		\nonumber
		\\
		\mathrm{ s.t. }
		&\quad
		\left| \mathbf{ z }(n)  \right| = 1, \; n=1, \ldots, N.
	\end{align}
	Enforcing $\mathbf{y}\triangleq\mathbf{z}\odot\mathbf{d}$ and $\mathbf{A}\triangleq\mathbf{R}^{-1}$, applying the constant-modulus constraint to $\mathbf{y}$, the problem \eqref{designSINR2} can be equivalently tackled via solving the CMCQP in \eqref{app2}.
	
	\subsubsection{MIMO-OFDM Communications Design}
	
	In the context of MIMO-OFDM downlink communications, $N$ subcarriers which construct $M$ resource blocks are employed to transfer symbols via a number of $N_{\mathrm{T}}$ antennas. Each resource block is allocated with user data/symbols by means of space-time block coding and inverse discrete Fourier transform (IDFT). Technically, the time-domain representation of the spatial data after unimodular precoding at transmission, denoted by $\mathbf{S} \in \mathbb{C}^{ N_{\mathrm{T}} \times N  } $, can be expressed as
	\begin{align}
		\mathbf{S}=\mathbf{W}^\mathrm{H}\mathbf{Y}\mathbf{D}\mathbf{F}
	\end{align}
	where $\mathbf{W}\in\mathbb{C}^{ M N_{\mathrm{T}}\times N_{\mathrm{T}}}$ is the beamforming matrix whose $M$ blocks correspond to the $M$ resource blocks of MIMO OFDM, $\mathbf{Y}\in\mathbb{C}^{MN_{\mathrm{T}}\times MN_{\mathrm{T}}}$ is the unimodular precoding matrix, $\mathbf{D}\in\mathbb{C}^{MN_{\mathrm{T}}\times N}$ is the transmit data matrix in the frequency domain, and $\mathbf{F} \in\mathbb{C}^{N\times N}$ is the IDFT matrix.
	
	Using the above model, one typical design to obtain the best PAPR for the time-domain transmission $\mathbf{S}$ \cite{PAPRR}, which equals $ \|\mathrm{vec}(\mathbf{S})\|_{\infty}^2 / \|\mathrm{vec}(\mathbf{S})\|^2 $, can be transformed as a form given by \eqref{app2}. For this application, we can select ${\mathbf{ y }}\triangleq\big[\mathbbm{d}^\mathrm{T}\{\mathbf{Y}\},e^{j\phi}\big]^\mathrm{T}$ with $\phi$ being an arbitrary phase value and
	\begin{align}
		{\mathbf{A}}^{(k)}\triangleq\begin{bmatrix}
			\mathbf{D}\mathbf{D}^\mathrm{H}\mathbf{W}\mathbf{W}^\mathrm{H}&\mathbf{a}^{(k)}\\
			\big(\mathbf{a}^{(k)}\big)^\mathrm{H}&\xi
		\end{bmatrix}
	\end{align}
	with $\mathbf{a}^{(k)}\triangleq\sqrt{\xi}((\mathbf{D}\mathbf{F})^\mathrm{T}\circ\mathbf{W}^\mathrm{H})^\mathrm{H}e^{j\cdot\arg((\mathbf{D}\mathbf{F})^\mathrm{T}\circ\mathbf{W}^\mathrm{H}\mathbbm{d}\{\mathbf{Y}^{(k)}\})}$ for optimization at the $k$-th iteration, and $\xi$ being the average transmit power per sample.

	\section{Simulations}
	
	In this section, we evaluate the performance of our proposed solutions to the CMCQP (i.e., Min/Max-CMCQP$^\text{I}$, Min/Max-CMCQP$^\text{II}$, and Min-CMCQP$^\text{WF}$ summarized in Algorithms~1, 2, and 3, respectively), wherein the conducted simulations are divided into three subsections. In these subsections, we show comprehensive evaluations in terms of different aspects for our proposed algorithms, including the step size, convergence speed, and multiple performance characteristics in the context of both CaseA1 and CaseA2. As regards each evaluation, the comparisons between our proposed algorithm and different counterparts are also presented. For the last evaluation in this section, we additionally show the performance comparisons between our proposed algorithm and other algorithms in specific applications addressed by the CMCQP. Throughout the simulations, we generate unit-modulus sequences with random phase values as initialization. We also randomly generate $\mathbf{A}$ with qualified structure that meets the requirement of CaseA1 or CaseA2. The same hardware and software configurations (i.e., PC with 3.20 GHz Intel Core i7-8700 CPU and 16 GB RAM running MATLAB R2019b) are used for comparisons. All the data are obtained over $50$ independent trials.

	\subsection{Step-Size Evaluations}
	
	In this subsection, we evaluate the accuracy of our proposed method for fast step-size determination. 
	Both accuracies of step size and search function for approximating their optima are tested, wherein the problem size is chosen as $N=30$.
	
	\begin{figure*}[t]
		\centering
		\subfloat[\label{st4d}] {\includegraphics[width=0.25\textwidth]{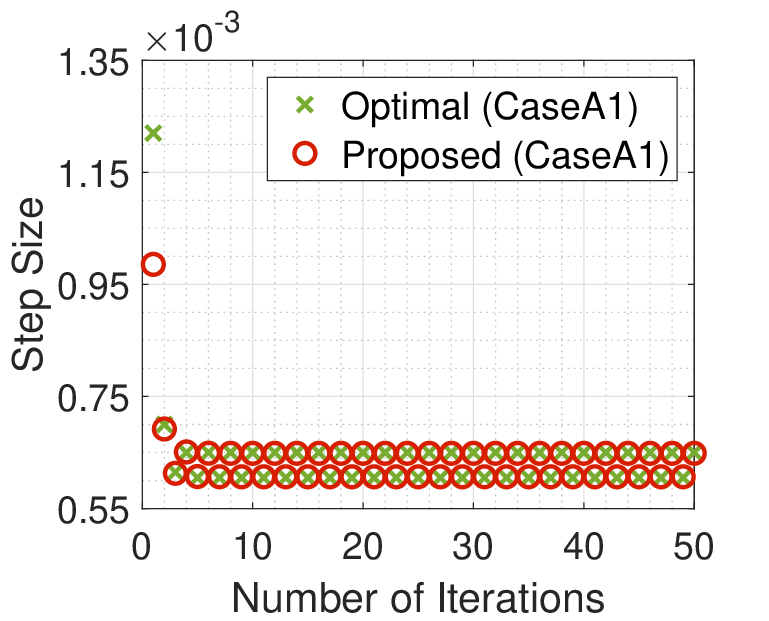}}
		\subfloat[\label{st4a}] {\includegraphics[width=0.25\textwidth]{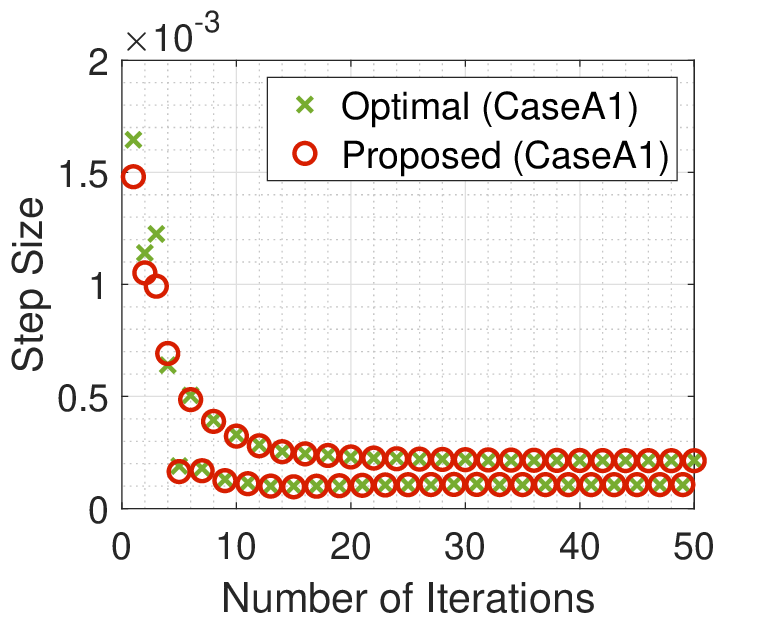}}
		\subfloat[\label{st2d}] {\includegraphics[width=0.25\textwidth]{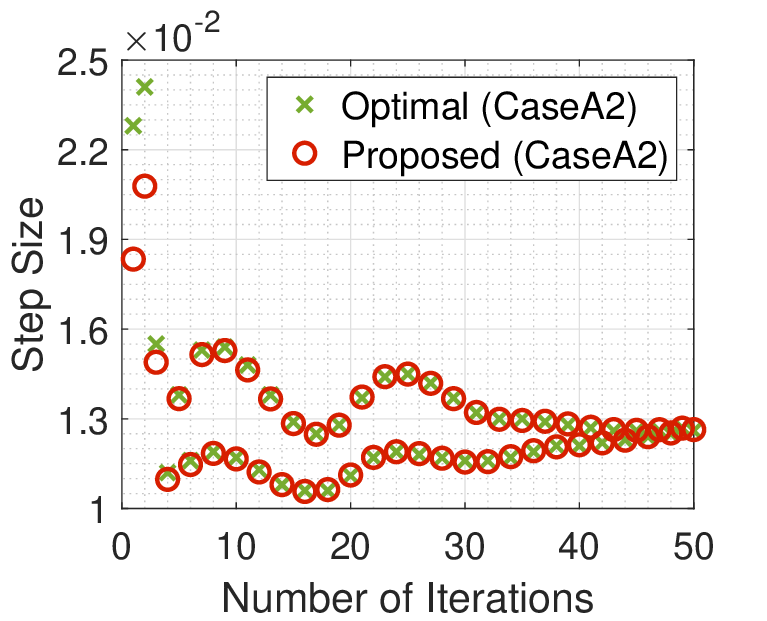}}
		\subfloat[\label{st2a}] {\includegraphics[width=0.25\textwidth]{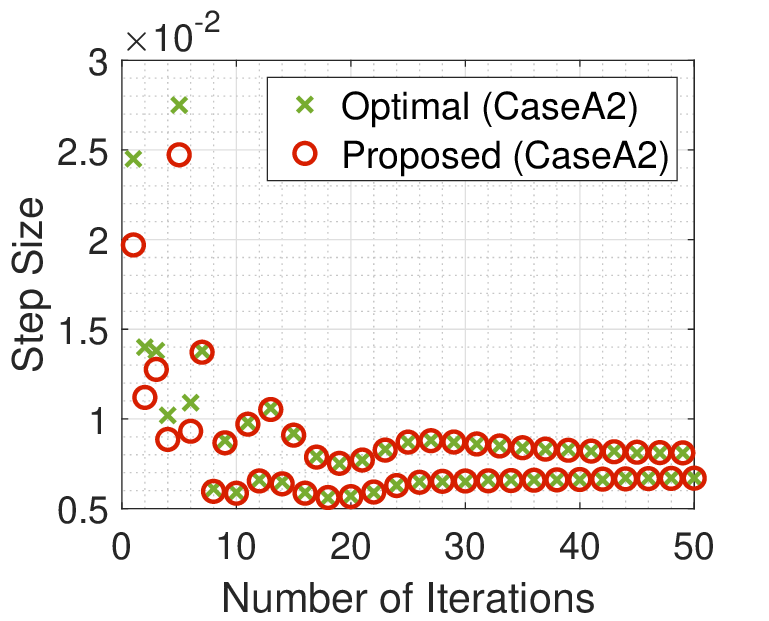}}
		\caption{Evaluations on the accuracies of step-size determinations to approximate their optima: (a) Minimization case for the CMCQP with CaseA1; (b) maximization case for the CMCQP with CaseA1; (c) minimization case for the CMCQP with CaseA2; and (d) maximization case for the CMCQP with CaseA2.}
		\label{st4}
	\end{figure*}
	\begin{figure*}[t]
		\centering
		\subfloat[\label{sf4d1}] {\includegraphics[width=0.25\textwidth]{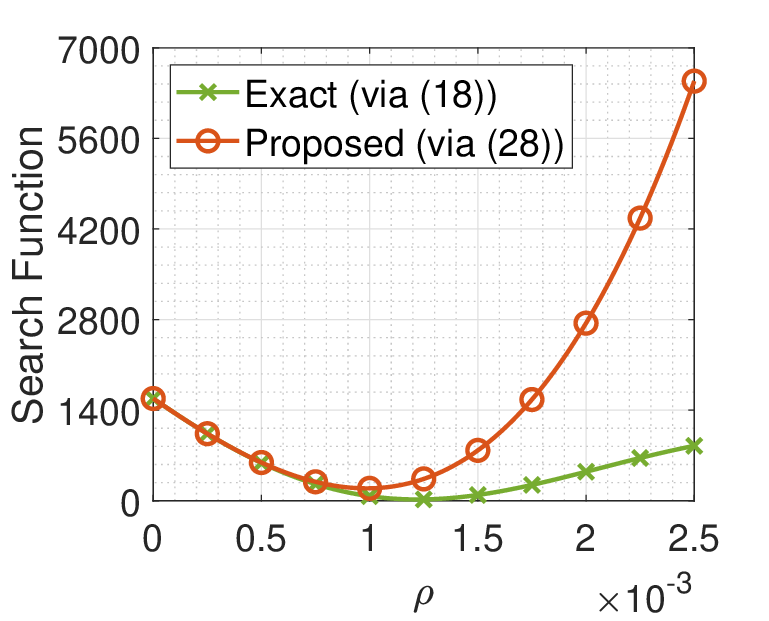}}
		\subfloat[\label{sf4d3}] {\includegraphics[width=0.25\textwidth]{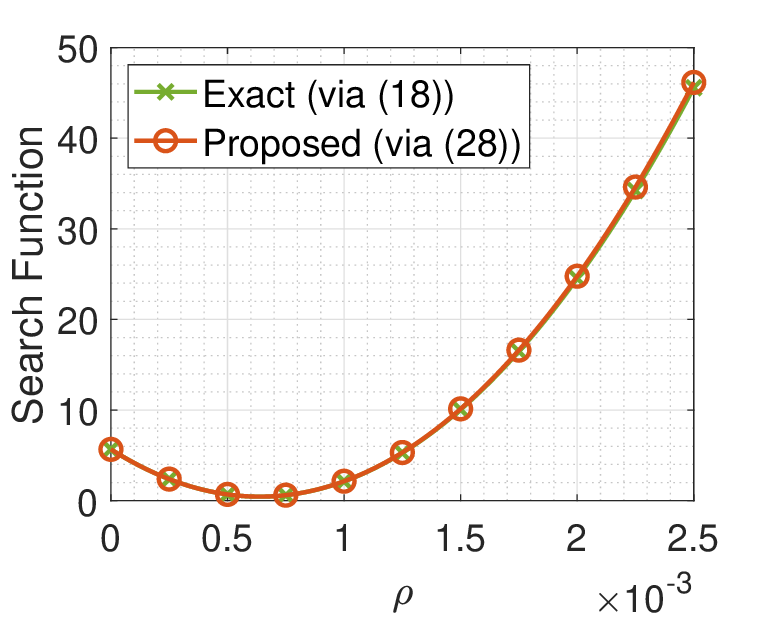}}
		\subfloat[\label{sf4a1}] {\includegraphics[width=0.25\textwidth]{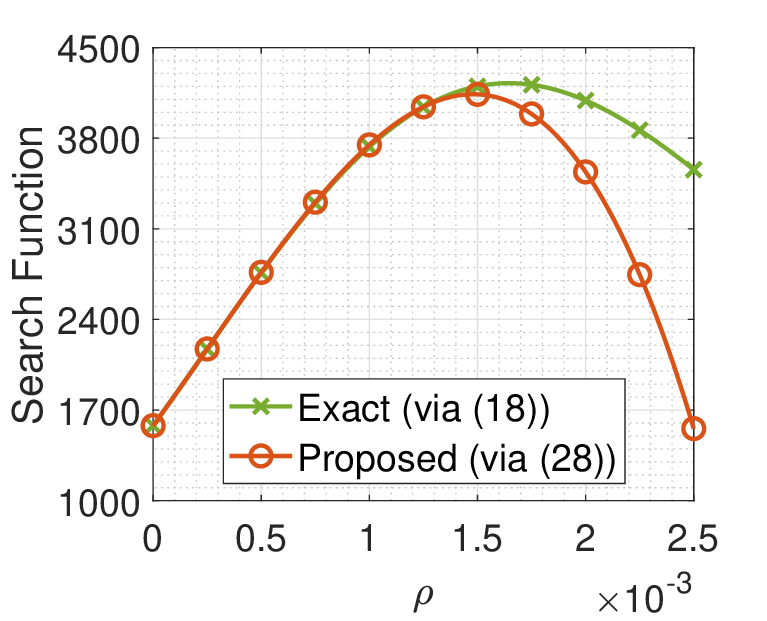}}
		\subfloat[\label{sf4a30}] {\includegraphics[width=0.25\textwidth]{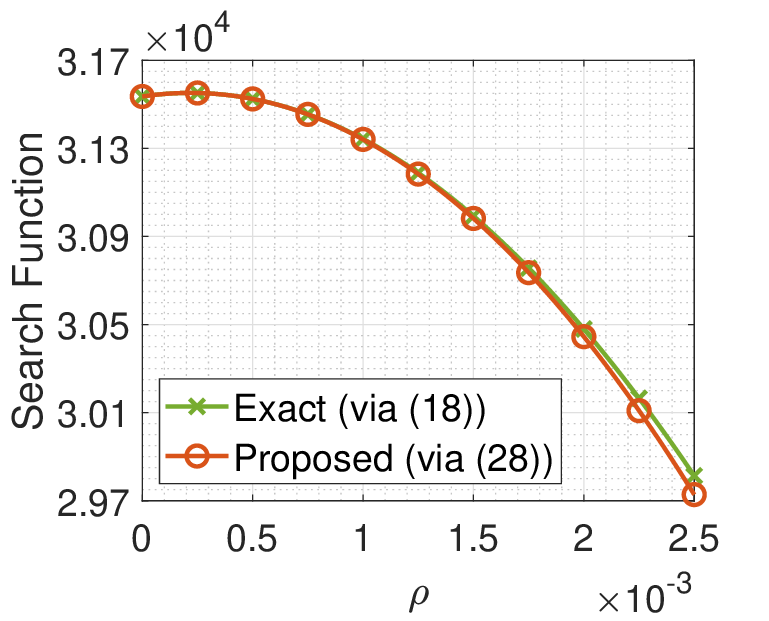}}\\
		\subfloat[\label{sf2d1}] {\includegraphics[width=0.25\textwidth]{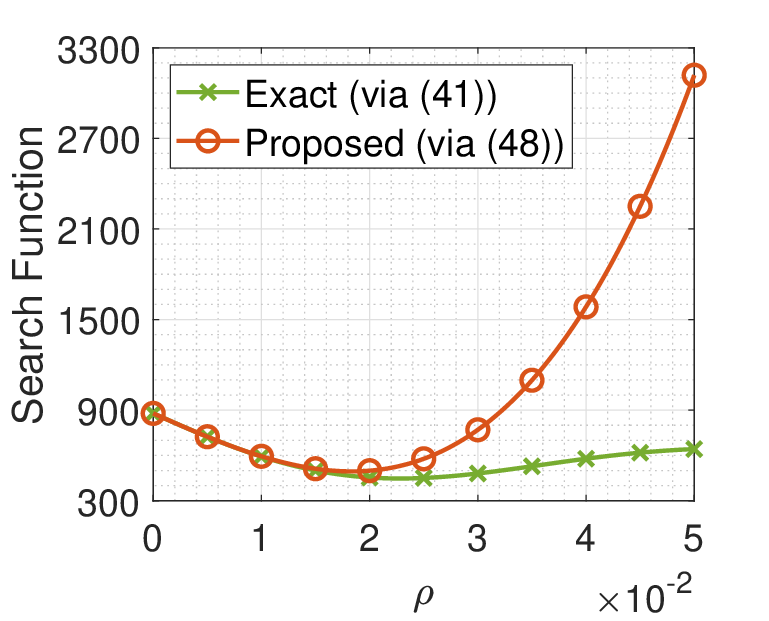}}
		\subfloat[\label{sf2d30}] {\includegraphics[width=0.25\textwidth]{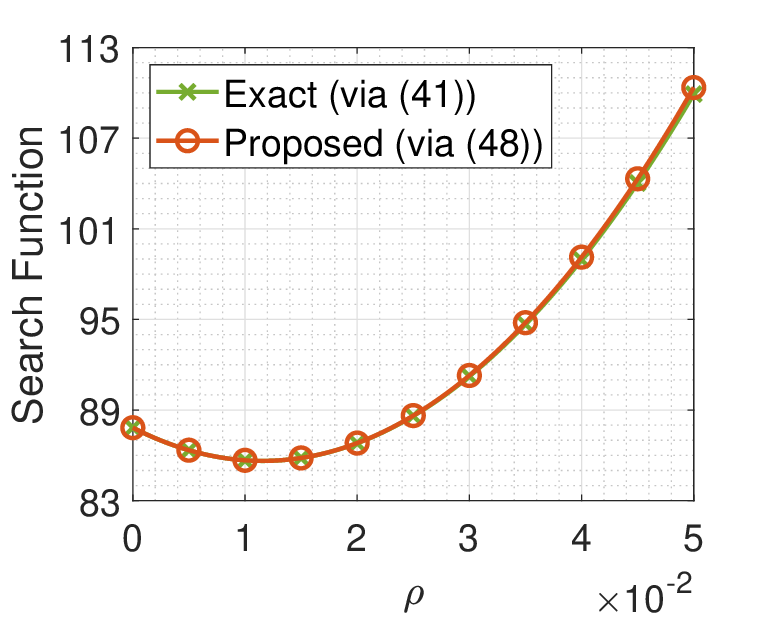}}
		\subfloat[\label{sf2a1}] {\includegraphics[width=0.25\textwidth]{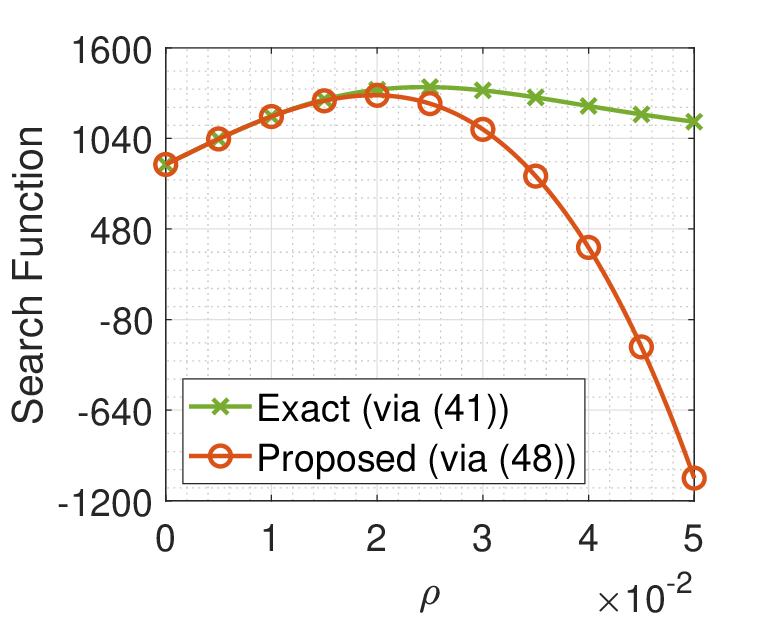}}
		\subfloat[\label{sf2a30}] {\includegraphics[width=0.25\textwidth]{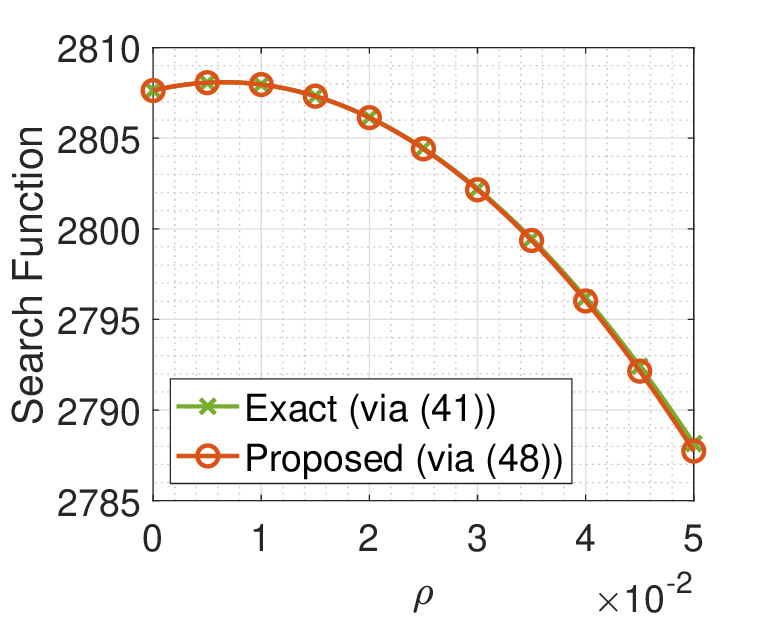}}
		\caption{Evaluations on the accuracies of search functions to approximate their corresponding optima: (a) Minimization case for CaseA1 ($k=1$); (b) minimization case for CaseA1 ($k=3$); (c) maximization case for CaseA1 ($k=1$); (d) maximization case for CaseA1 ($k=30$); (e) minimization case for CaseA2 ($k=1$); (f) minimization case for CaseA2 ($k=30$); (g) maximization case for CaseA2 ($k=1$); and (h) maximization case for CaseA2 ($k=30$).}
		\label{objf}
	\end{figure*}
	
	\subsubsection{Step-Size Accuracy}
	
	The results for evaluating the step-size accuracy of our proposed method along $50$ iterations are shown in Fig.~\ref{st4}, wherein the first two sub figures (Figs.~\ref{st4}\subref{st4d} and \ref{st4}\subref{st4a}) correspond to CaseA1, and the last two sub figures (Figs.~\ref{st4}\subref{st2d} and \ref{st4}\subref{st2a}) correspond to CaseA2. The optimum step sizes presented here are obtained by the exact line search \cite{CVX}, and the metric to evaluate the accuracy of our step-size determination is $100\%$ minus its relative error to the optimal value. It can be seen from Figs.~\ref{st4}(a) and \ref{st4}(b) that our proposed fast step-size determination method shows good accuracy of approaching the optimum step size in CaseA1, which has achieved a high level nearly up to $100\%$ after the first one and four iterations for the minimization and maximization cases, respectively.
		The biggest difference between our determined and the optimum step size values within the first four iterations is about $4\times10^{-4}$ (see Fig.~\ref{st4}(a)), which almost has no effect on the overall performance of algorithm iterations because its keeps the monotonous trend of the objective value. Note that the small difference for fast step-size determination at the very early stage is normally unavoidable because of the random guess for initialization. The large gradients of the initial points degrade the performance of our Taylor-expansion based approximation for the step-size search function. However, this degradation can be avoided after iterations when the norm of gradient gradually becomes small.
		Using our proposed algorithms Min-CMCQP$^\text{I}$ and Max-CMCQP$^\text{I}$ with the step-size correction strategy given by \eqref{stepcorrect}, the determinations on step sizes converge to stable and accurate points with fast speeds.
		
		For the step-size accuracy tested in CaseA2, it can be seen from Figs.~\ref{st4}\subref{st2d} and \ref{st4}\subref{st2a} that similar trends on the step-size determinations occur, wherein the step-size accuracy has also reached a high level close to $100\%$ after at most $6$ iterations. Moreover, the biggest gap between our determined and the optimum step sizes is about $0.0048$. Specifically, the approximation gaps of our proposed step-size determinations for minimization and maximization cases are only visible before $3$ and $6$ iterations, respectively.
		Relatively, the gaps before precise approximation for the latter case are larger than those for the former. The minor difference between the results for CaseA1 and CaseA2 only lies in the fast obtained stable points.

	\subsubsection{Search-Function Accuracy} 
	
	The accuracies of our proposed search functions \eqref{Opt4} and \eqref{Opt2} to respectively approximate their optima \eqref{Opt4e} and \eqref{Opt2e} are shown in Fig.~\ref{objf}, wherein the values of search functions versus step sizes are plotted under both minimization and maximization cases. The first and last four sub figures of Fig.~\ref{objf} correspond to CaseA1 and CaseA2, where the tested step sizes range from $0$ to $0.05$ and $0$ to $0.0025$, respectively. In each case, we show the results obtained at the initial and stable stages of iterations. 
	
	It can be seen from Fig.~\ref{objf} that our proposed methods show high accuracy of approximating the optimum after arriving at stable stages of iterations (see Figs.~\ref{objf}\subref{sf4d3}, \ref{objf}\subref{sf4a30}, \ref{objf}\subref{sf2d30}, and \ref{objf}\subref{sf2a30}). After $3$ and $30$ iterations, our proposed search functions \eqref{Opt4} and \eqref{Opt2} fully overlap \eqref{Opt4e} and \eqref{Opt2e} for both CaseA1 and CaseA2, respectively. At the initial stage of iterations (see $k=1$ in Figs.~\ref{objf}\subref{sf4d1}, \ref{objf}\subref{sf4a1}, \ref{objf}\subref{sf2d1}, and \ref{objf}\subref{sf2a1}), there exist some biases between the approximated and optimal search functions when the tested step size becomes relatively large, which are also unavoidable because of random guesses at initialization. However, such biases are corrected as our proposed methods transit to stable states after a certain small number of iterations. Indeed, the approximated search functions at the initial stage of iterations already lead to good step-size determinations since the optimal solutions are near to their local minima/maxima. In other words, the optimal step sizes are only determined by the local extremes of search functions, which are well approximated by the corresponding extreme points of our proposed search functions even at the beginning of iterations.

	\subsection{Convergence Evaluations}
	
	\begin{figure*}[t]
		\centering
		\subfloat[\label{con4d}] {\includegraphics[width=0.25\textwidth]{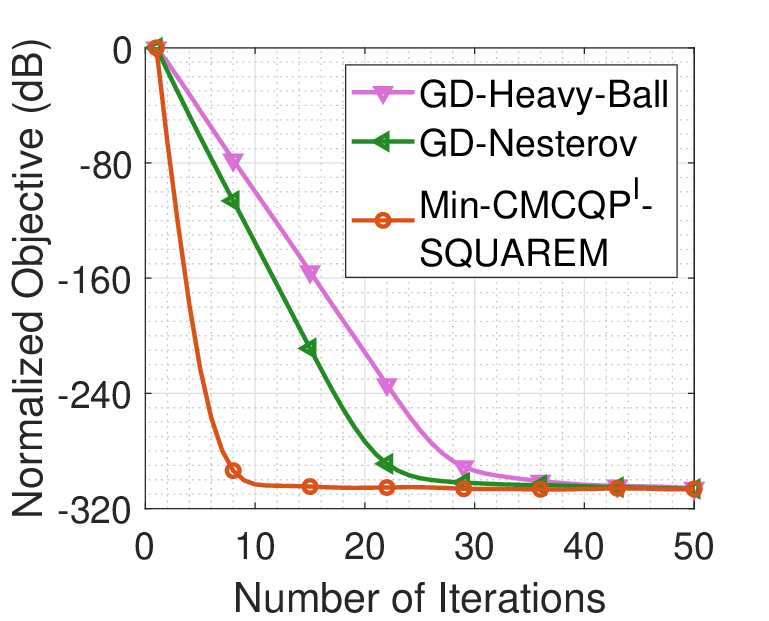}}
		\subfloat[\label{con4a}] {\includegraphics[width=0.25\textwidth]{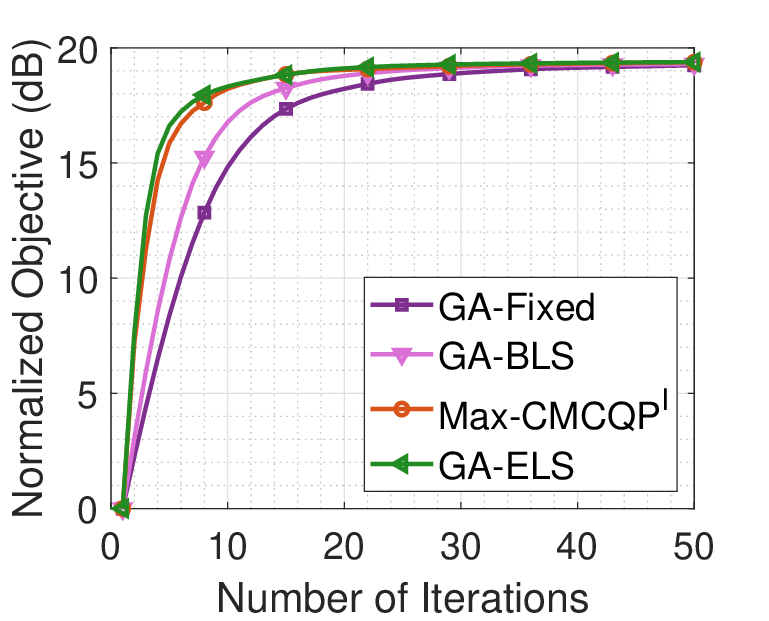}}
		\subfloat[\label{con2d}] {\includegraphics[width=0.25\textwidth]{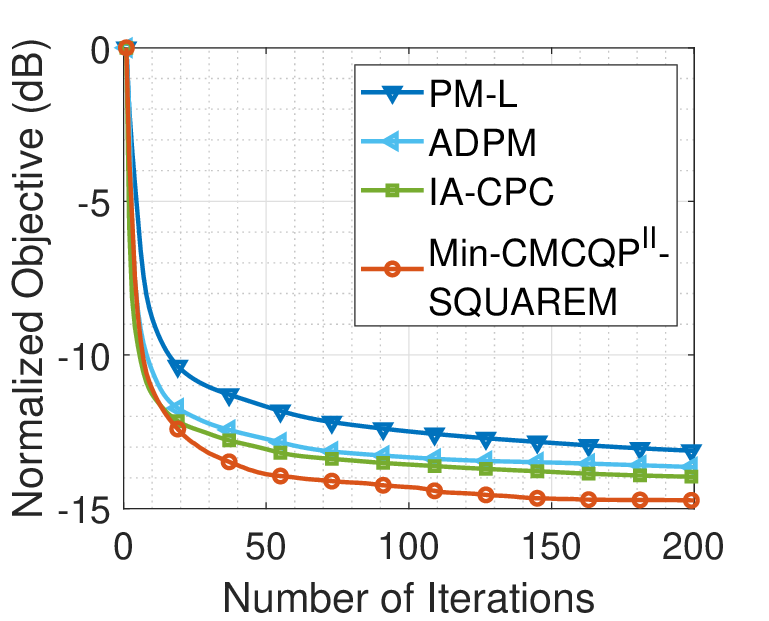}}
		{\color{red} \subfloat[\label{con2a}] {\includegraphics[width=0.25\textwidth]{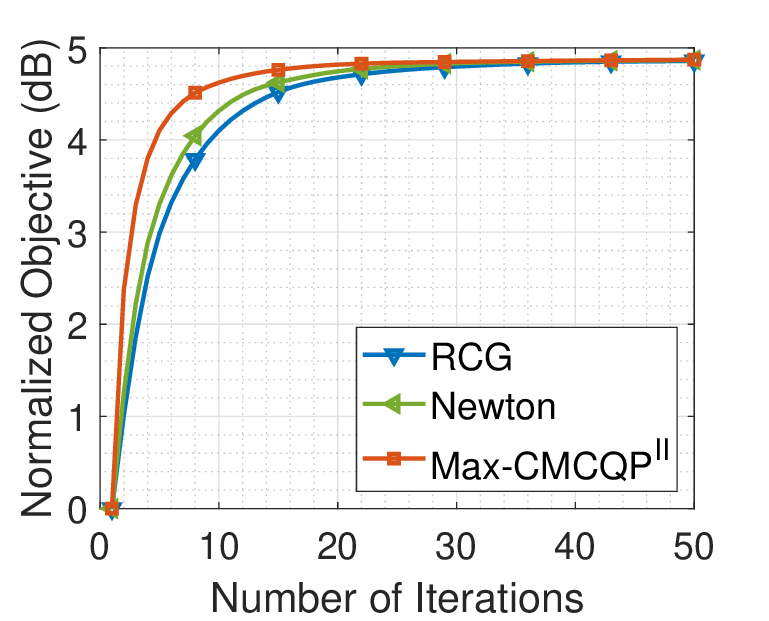}}
			\caption{Evaluations on the convergence speeds of algorithms versus iterations: (a) Minimization case for CaseA1 ($N=1000$); (b) maximization case for CaseA1 ($N=30$); (c) minimization case for CaseA2 ($N=1000$); and (d) maximization case for CaseA2 ($N=30$).}
			\label{conv}	}
	\end{figure*}
	
	In this subsection, we evaluate the convergence speeds of our proposed algorithms versus iterations for both CaseA1 and CaseA2. We conduct multiple types of comparisons between our methods and existing algorithms that are able to address CaseA1 or CaseA2. Specifically,  in the minimization case of CaseA1, we test different acceleration schemes. The adopted algorithms include GD with Heavy-Ball and Nesterov accelerations (named as `GD-Heavy-Ball' and `GD-Nesterov', respectively), and our proposed Min-CMCQP$^{\mathrm{I}}$ with SQUAREM acceleration \cite{accEM}. In the case of maximization, we investigate different step-size determination systems. The algorithms chosen here are Max-CMCQP$^{\mathrm{I}}$ (without acceleration) and GA with a fixed value, backtracking line search, and exhaustive line search for step sizes (named as `GA-Fixed', `GA-BLS', and `GA-ELS', respectively). 
		
		Regarding CaseA2, we compare the SQUAREM accelerated Min-CMCQP$^{\mathrm{II}}$ with the power method-like algorithm of \cite{UQPstoica} (named as `PM-L') and the algorithms of \cite{UQPCui} and \cite{UQPCui2} (named as `IA-CPC' and `ADPM', respectively) in the minimization case. The key steps of them are as follows. PM-L implements a direct constant-modulus projection on the complex gradient. IA-CPC optimizes the objective element-wise through coordinate ascent. ADPM includes a binary decision for penalty determinations in the standard alternating direction method of multipliers. For the maximization case, we study various gradient-based search directions. The tested algorithms are our proposed Max-CMCQP$^{\mathrm{II}}$ (without acceleration), Riemannian conjugate gradient method (named as `RCG'), and Newton's method (named as `Newton'). Throughout the simulations here, problem sizes $N=30$ and $1000$ are selected for minimization and maximization cases, respectively.
	
	The evaluation results on convergence speeds of algorithms are shown in Fig.~\ref{conv}. The first and last halves therein correspond to CaseA1 and CaseA2, respectively, each of which includes two sub figures for both minimization and maximization cases. 
	It can be seen from Fig.~\ref{conv}\subref{con4d} that our proposed algorithm Min-CMCQP$^\text{I}$ with SQUAREM acceleration shows a much faster convergence speed than those of GD-Heavy-Ball and GD-Nesterov. Specifically, our proposed algorithm arrive to stable points after around $15$ iterations for the minimization case in CaseA1, while its counterparts spend more than $40$ iterations to achieve the same goal of being stable. As for the maximization case, our proposed algorithm Max-CMCQP$^\text{I}$ shows the most closed convergence performance to the bound of GA regime, i.e., GA-ELS. The improvements of Max-CMCQP$^\text{I}$ over GA-Fixed and GA-BLS on calculating normalized objective values of the CMCQP observed at the $10$-th iteration are $4.78$ and $2.36$~dBs, respectively.

	When it comes to CaseA2, it can be seen from Figs.~\ref{conv}\subref{con2d} and \ref{conv}\subref{con2a} that our proposed algorithms  Min-CMCQP$^\text{II}$ and Max-CMCQP$^\text{II}$ show the best convergence speeds over all the other tested algorithms. After $50$ iterations, our proposed algorithm Min-CMCQP$^\text{II}$ achieves a normalized objective function valued $-13.88$ dB in the minimization case, while the values of the normalized objective function obtained by algorithms IA-CPC, ADPM, and PM-L are $-13.07$, $-12.73$, and $-11.68$ dBs, respectively. As for the maximization case, after $10$ iterations, the values of normalized objective function obtained by Max-CMCQP$^\text{II}$, Newton, and RCG reach $4.62$, $4.32$, and $4.10$~dBs, respectively, which respectively account for $94.87\%$, $88.71\%$, and $84.19\%$ of the maximally achievable value equaling $4.87$~dB. The reason for the slow convergence speeds of Newton and RCG lies in their poor step-size determination systems. It is of significant difficulty to fast estimate their exact optimal step sizes, so that we adopt a suboptimal fixed step size for each, which therefore presents undesired convergence performances.

	\subsection{Performance-Characteristic Evaluations}
	
	In this subsection, we evaluate the performance characteristics of our proposed algorithms (the SQUAREM accelerated versions) versus problem sizes, and also compare with existing algorithms in terms of multiple characteristics including i) the average time consumption (in seconds); ii) the average number of iterations; and iii) the minimum, maximum, and averaged values of objective function obtained after convergence.
	The simulations here are divided into two parts. One conducts performance evaluations and algorithm comparisons for the CMCQP in mathematics, wherein the matrix $\mathbf{A}$ is randomly generated. For completeness, the other compares algorithms in the context of applications, wherein the matrix $\mathbf{A}$ is chosen based on specific tasks addressed by the CMCQP. 
	Throughout simulations, the stopping criterion for comparison is defined as the absolute difference on the values of objective function between two neighboring iterations normalized by the value at initialization, whose tolerance is set to be $10^{-9}$.
	
	\subsubsection{Performance Characteristics in Mathematics}
	
	For pure mathematical experiments, we choose CaseA2 as the representative since it allows for more available methods to conduct comparisons than those addressing CaseA1. We first investigate the minimization case and compare our proposed Min-CMCQP$^\text{II}$ with PM-L, IA-CPC, and ADPM, wherein the tested problem sizes are chosen from the set $N\in\{100,200,300,400,500\}$. The comparison results are shown in Table~1. It can be seen that our proposed algorithm Min-CMCQP$^\text{II}$ behaves the best in terms of all performance characteristics for all problem sizes that are tested. It especially shows considerable improvements in terms of time consumption and number of iterations. For instance, it has reduced the time consumption at $N=500$ by around $11$, $14$, and $53$ times ($0.23$ versus $2.62$, $3.28$, and $12.08$ seconds) compared to PM-L, IA-CPC, and ADPM, respectively. The algorithms IA-CPC and ADPM cause high time consumption due to the algorithmic frameworks used therein. Our proposed algorithm also obtains slightly better improvements on the normalized objective value after convergence.
		
		\begin{table*}[t]
			\setlength{\extrarowheight}{2pt}
			\centering
			\abovecaptionskip = 7pt
			\caption{Performance Comparisons Versus Different Problem Sizes for the Minimization Case of the CMCQP in CaseA2.}
			\vspace*{-4pt}
			\begin{adjustbox}{max width=\textwidth}
				\begin{threeparttable}
					\centering
					\setlength\tabcolsep{3.5pt}
					\begin{tabular}{|c|c|c|c|c|c|c|c|c|c|c|c|c|c|c|c|c|c|c|c|c|c}
						\hline
						\multicolumn{1}{|c|}{\multirow{2}{*}{}} 	& \multicolumn{4}{c|} {$ N=100 $}	& \multicolumn{4}{c|} {$ N=200 $}& \multicolumn{4}{c|} {$ N=300 $} 	& \multicolumn{4}{c|} {$ N=400 $} 	& \multicolumn{4}{c|} {$ N=500 $}
						\\ \cline{2-21}
						&  Time  & Iter.\tnote{\emph{a}}\,  &  Min.\tnote{\emph{b}}\,      &  Ave.\tnote{\emph{c}}\,     &  Time  & Iter. &Min. & Ave. &  Time  & Iter. &Min. & Ave.         &  Time     & Iter. &Min.  &  Ave.  & Time        &  Iter.                   &Min.    & Ave.
						\\	\hline
						ADPM	  & 0.22  &	1136 &-13.72	&	-13.65	&	1.67		&  1994 &-14.09	&	-14.05&4.79&1086&-14.45&-13.95&7.88&1049&-14.68&-14.40	&	12.08	& 958 	&-14.81&-14.45
						\\ \hline
						IA-CPC		&	0.11	&	324	&-13.75	&	-13.67		&	0.47	&	547	&-14.13 &	-14.05&1.27&865&-14.61&-14.35&2.13&988&-14.69&-14.35	&	3.28	&1072	&-14.77&	-14.43
						\\ \hline
						PM-L			& 0.073  &	2836 &-13.76	&	-13.61	&	0.28		& 5724 &-14.07&	-13.88&0.63&7740&-14.55&-14.30&1.31&9179&-14.72&-14.40	&	2.62	&	11675		&-14.73& -14.43 
						\\ \hline
						Min-CMCQP$^\text{II}$		& \bf{0.0096}	&	\bf{135}	&\bf{-13.78}	&	\bf{-13.74}	&	\bf{0.030}	& \bf{210}	&\bf{-14.13}&	\bf{-14.06}&\bf{0.073}&\bf{299}&\bf{-14.63}&\bf{-14.42}&\bf{0.16}&\bf{424}&\bf{-14.75}&\bf{-14.45}	&	\bf{0.23} 	&	\bf{436}	&\bf{-14.82}& \bf{-14.46} 
						\\ \hline
					\end{tabular}
				\end{threeparttable}
			\end{adjustbox}
			\begin{tablenotes}
				\vspace*{-8pt}
				{\fontsize{7.5}{8}\selectfont
					\item[\emph{a}]Iter.: Average number of conducted iterations.
					\item[\emph{b}]Min.: Minimum normalized objective value (in dB).
					\item[\emph{c}]Ave.: Average normalized objective value (in dB).}
			\end{tablenotes}
		\end{table*}

		\begin{table*}[t]
			\setlength{\extrarowheight}{2pt}
			\centering
			\abovecaptionskip = 7pt
				\caption{Performance Comparisons Versus Large-Scale Problem Sizes for the Maximization Case of the CMCQP in CaseA2.}
				\vspace*{-4pt}
				\begin{adjustbox}{max width=\textwidth}
					\begin{threeparttable}
						\centering
						\setlength\tabcolsep{5pt}
						\begin{tabular}{|c|c|c|c|c|c|c|c|c|c|c|c|c|c|c|c|c|c|c|c|c|c}
							\hline
							\multicolumn{1}{|c|}{\multirow{2}{*}{}} 	& \multicolumn{5}{c|} {$ N=1024 $}	& \multicolumn{5}{c|} {$ N=2048 $} & \multicolumn{5}{c|} {$ N=4096 $}& \multicolumn{5}{c|} {$ N=8192 $}	
							\\ \cline{2-21}
							&  Time & Iter.\tnote{\emph{a}}\,  &  Max.\tnote{\emph{b}}\,      &  Ave.\tnote{\emph{c}}\,      &  Min.\tnote{\emph{d}}\,  & Time & Iter. & Max. & Ave. & Min. & Time & Iter. & Max. & Ave. & Min. & Time & Iter. & Max. & Ave. & Min.
							\\	\hline
							Newton	  & 87.23 & 390 & 5.42 & 5.35 & 5.26  & 766.46 & 573 & 5.50 & 5.43 & 5.31 & 1827.67 & 1104 & 5.61 & 5.50 & 5.43 & hours & - & - & - & -
							\\ \hline
							RCG		&  2.01 & 425 & 5.41 & 5.36 & 5.24 & 11.33 & 716 & 5.51 & 5.43 & 5.30 & 73.93 & 1365 & 5.61 & 5.51 & 5.42 & 678.83 & 3399 & 5.66 & 5.64 & 5.63
							\\ \hline
							Max-CMCQP$^\text{II}$	& \bf{1.27} & \bf{259} & \bf{5.43} &\bf{5.36} & \bf{5.26} & \bf{8.29} & \bf{377} & \bf{5.51} & \bf{5.44} & \bf{5.32} & \bf{51.75} & \bf{610} & \bf{5.62} & \bf{5.51} & \bf{5.43} & \bf{505.65} & \bf{1564} & \bf{5.67} & \bf{5.65} & \bf{5.63}
							\\ \hline
						\end{tabular}
					\end{threeparttable}
				\end{adjustbox}
				\begin{tablenotes}
					\vspace*{-8pt}
					{\fontsize{7.5}{8}\selectfont
						\item[\emph{a}]Iter.: Average number of conducted iterations.
						\item[\emph{b}]Max.: Maximum normalized objective value (in dB).
						\item[\emph{c}]Ave.: Average normalized objective value (in dB).\\
						\item[\emph{d}]Min.: Minimum normalized objective value (in dB).}
			\end{tablenotes}
		\end{table*}
		
		We then study the maximization case and compare our proposed algorithm with RCG and Newton (all accelerated by SQUAREM), wherein large-scale problem sizes $N\in\{2^{10}, 2^{11}, 2^{12}, 2^{13}\}$ are tested. The corresponding results are shown in Table~2. It can be seen that our proposed Max-CMCQP$^\text{II}$ also outperforms all the other tested algorithms in terms of all the performance characteristics. The improvement over time and iteration consumption is also significant. When the problem size is as large as $8192$, our proposed algorithm spends $505.65$ seconds with $1564$ iterations for convergence. In contrast, RCG takes $678.83$ seconds with $3399$ iterations, while Newton requires hours or even more time to converge. In general, Newton and RCG show heavy computational burdens because of the high complexity to calculate the Hessian matrix and the repetitive projections onto constant-modulus manifolds, respectively. It can also be seen from Table~2 that all the tested algorithms are less sensitive to the initialization. Among $50$ independent trials with different initial points, the biggest gap between the maximum and minimum normalized objective values is $0.21$ dB (see the results of RCG when $N=2048$ for evidence).
	
	\subsubsection{Performance Characteristics in Applications}
	
	We also conduct performance evaluations on the aforementioned performance characteristics of algorithms in the context of applications. 
	Two applications are chosen to verify the effectiveness of our proposed algorithms for CaseA1 and CaseA2, respectively. The first application is about the unimodular waveform/code design, and the other is about the optimum detection design.
	
	\begin{table*}[t]
		\setlength{\extrarowheight}{2pt}
		\centering
		\abovecaptionskip = 7pt
		\caption{WISL Performance Comparisons Versus Different Numbers of Waveforms With Fixed Code Length.}
		\vspace*{-4pt}
		\begin{adjustbox}{max width=\textwidth}
			\begin{threeparttable}
				\centering
				\setlength\tabcolsep{4.1pt}
				\begin{tabular}{|c|c|c|c|c|c|c|c|c|c|c|c|c|c|c|c|c|c|c|c|c|c}
					\hline
					\multicolumn{1}{|c|}{\multirow{2}{*}{}} 	& \multicolumn{4}{c|} {$ P=128,M=3 $}	& \multicolumn{4}{c|} {$ P=128,M=4 $} & \multicolumn{4}{c|} {$ P=128,M=5 $}& \multicolumn{4}{c|} {$  P=128,M=6 $}& \multicolumn{4}{c|} {$  P=128,M=7 $}	 	\\ \cline{2-21}
					&  Time  & Iter.\tnote{\emph{a}}\,                       &  Min.\tnote{\emph{b}}\,  &   Ave.\tnote{\emph{c}}\,          &  Time  & Iter. & Min. &Ave.  &  Time  & Iter. & Min. &Ave.               &  Time                      & Iter. & Min. &Ave. &  Time    & Iter. & Min. &Ave.  \\	\hline
					WeCAN	  &59.23 &2987 &22.24 &23.12 &161.29 &5406 &27.11 &27.96 &247.35 &7963 & 34.77 & 35.29 & 283.64 & 7539 &40.86 &41.21&422.06&9962&45.63&45.79\\ \hline
					MM-WeCorr	&	11.50	&	272		&	-22.23		&	-13.69	&70.47 &998&22.35 &23.18 &90.48 &1027 &33.44 &33.97 &192.96 &1214 &40.29 &40.68&273.24&1181&44.92&45.18	\\ \hline
					WISLNew		& 4.44&	181	&	-28.82	&	-22.48 & 34.07 & 857 &22.18 & 23.02 &55.16 &866 &33.38 &33.85 &94.76 & 1025 &40.22 &40.61&164.35&1315&44.74&45.16	    \\ \hline
					Min-CMCQP$^\text{WF}$	&\bf{ 0.21}  &	\bf{131}		&	\bf{-38.91}	&	\bf{-25.53} &\bf{1.24} & \bf{569} & \bf{21.83} &\bf{22.91} &\bf{1.39} &\bf{422} &\bf{33.14} &\bf{33.81} &\bf{2.71} &\bf{529} &\bf{40.12} &\bf{40.58}&\bf{3.92} &\bf{695} &\bf{44.72} &\bf{45.08}	 \\ \hline
				\end{tabular}
			\end{threeparttable}
		\end{adjustbox}
		\begin{tablenotes}
			\vspace*{-8pt}
			{\fontsize{7.5}{8}\selectfont
				\item[\emph{a}]Iter.: Average number of conducted iterations.
				\item[\emph{b}]Min.: Minimum WISL value (in dB).
				\item[\emph{c}]Ave.: Average WISL value (in dB).}
		\end{tablenotes}
	\end{table*}
	
	\begin{table*}[t]
		\setlength{\extrarowheight}{2pt}
		\centering
		\abovecaptionskip = 7pt
		\caption{SNR Performance Comparisons Versus Different Code Lengths.}
		\vspace*{-4pt}
		\begin{adjustbox}{max width=\textwidth}
			\begin{threeparttable}
				\centering
				\setlength\tabcolsep{3.5pt}
				\begin{tabular}{|c|c|c|c|c|c|c|c|c|c|c|c|c|c|c|c|c|c|c|c|c|c}
					\hline
					\multicolumn{1}{|c|}{\multirow{2}{*}{}} & \multicolumn{4}{c|} {$ N=64 $}	& \multicolumn{4}{c|} {$ N=128 $} & \multicolumn{4}{c|} {$ N=256 $}& \multicolumn{4}{c|} {$ N=512 $}	& \multicolumn{4}{c|} {$ N=1024 $} 	\\ \cline{2-21}
					&  Time  & Iter.\tnote{\emph{a}}\,                       &  Max.\tnote{\emph{b}}\, & Ave.\tnote{\emph{c}}\,                    &  Time  & Iter. & Max. &Ave.                 &  Time                      & Iter. & Max. &Ave.  & Time                        &  Iter.                      & Max. &Ave. & Time                        &  Iter.                      & Max. &Ave. \\	\hline
					ADPM & 1.23  &	7402	&	27.54&	27.54	&	15.61		&  23877	&	30.58&	30.58 &150.79 &54817 &33.61 &33.61 &2645.18&164144&36.63 &36.63	&	hours	& - 	&- &-\\ \hline
					IA-CPC 	&	0.41	&	1894		&	27.54&	27.54		&	2.51	&	5251	&	30.58&	30.58 &21.56 &17586 &33.61&33.61&138.36&68844&36.63	&36.63 &	1906.51	&123063	&	39.64 &	39.64		\\ \hline
					PM-L	& 0.11	&	6888	&	27.54&	27.54	&	0.71		& 21038 &	30.58&	30.58 &4.58&57644&33.61&33.61&27.95&139758&36.63&36.63	&	230.09	&	126632		& 39.64   &	39.64 \\ \hline
					Max-CMCQP$^\text{II}$ & \bf{0.021}  & \bf{325}	&	27.54&	27.54	&	\bf{0.084}	& \bf{824}	&	30.58&	30.58 &\bf{0.49}&\bf{2215}&33.61&33.61&\bf{2.81}&\bf{5589}&36.63&36.63	&	\bf{46.04} 	&	\bf{9730}	& 39.64 &	39.64\\ \hline
				\end{tabular}
			\end{threeparttable}
		\end{adjustbox}
		\begin{tablenotes}
			\vspace*{-8pt}
			{\fontsize{7.5}{8}\selectfont
				\item[\emph{a}]Iter.: Average number of conducted iterations.
				\item[\emph{b}]Max.: Maximum SNR value (in dB).
				\item[\emph{c}]Ave.: Average SNR value (in dB).}
		\end{tablenotes}
	\end{table*}
	
	\emph{Application~I---Multi-unimodular Waveform design}:
	The application of designing multiple unimodular waveforms with good correlation properties is investigated, where the WISL minimization based design addressed by the minimization case of the CMCQP for CaseA1 is tested. Throughout the simulations here, the number of waveforms is chosen from the set $M\in\{3,4,5,6,7\}$, and the code length of waveforms is fixed as $P=128$. The ISL controlling weights are set to be $\gamma_p=1,\,p=-19,\cdots,19$, while the others are all zeros. For this evaluation, we compare our proposed algorithm Min-CMCQP$^\text{WF}$ in Algorithm 3 with algorithms WeCAN \cite{HeMIMO09}, MM-WeCorr \cite{SongWFMaMi16}, and WISLNew \cite{LiWF18}. 
	
	The corresponding results are shown in Table~3. It can be seen from Table~3 that our proposed algorithm outperform all the other algorithms in terms of all performance characteristics for all the tested cases. The biggest gaps on the minimum and averaged WISL values obtained by our proposed Min-CMCQP$^\text{WF}$ and the second best WISLNew have reached $10.09$ dB and $3.05$ dB, respectively (see the case associated with $P=128$ and $M=3$). In particular, our proposed algorithm Min-CMCQP$^\text{WF}$ shows significant performance improvements in the aspects of time consumption and number of iterations. For example, it costs $3.92$ seconds associated with $695$ iterations to converge to the preset tolerance for $P=128$ and $M=7$, while WISLNew, MM-WeCorr, and WeCAN require $164.35$, $273.24$, and $422.06$ seconds associated with $1315$, $1181$, and $9962$ iterations, respectively.  Generally, the other tested algorithms spend a ten/hundredfold increase on the computational time compared to our proposed algorithm.
	
	\emph{Application II---Optimum Detection Design}. 
	The application of optimum detection design is studied, where the SNR maximization design addressed by the maximization case of the CMCQP for CaseA2 is tested. For this evaluation, we compare our proposed algorithm with publicly available algorithms. 
	For this application, the target is set to have a normalized Doppler frequency equaling $0.2$, and the disturbance matrix is set to have the same form as that in \cite{DeMaioPCode09} (i.e., the $(n,n')$-th element equals $ 0.8^{|n-n'|}$).
	Multiple code lengths given by $N=\{2^6,2^7,2^8,2^9,2^{10}\}$ are tested. 
	The corresponding comparison results are shown in Table~4. 
	It can be seen that all algorithms achieve the same maximal and average SNR values for all cases that are tested, but our proposed method consumes the least number of iterations and amount of time. 
	For example, in the case of $N = 1024$, the ADPM costs several hours to converge to tolerance, while the IA-CPC and PM-L cost around $1907$ and $230$ seconds over $123$ and $126$ thousands of iterations, respectively. Our proposed Max-CMCQP$^\text{II}$ only spends $46.04$ seconds via $9730$ iterations.
	
	\section{Conclusion}
	
	In this paper, we have studied the minimization/maximization on a multi-order complex quadratic form with constant-modulus constraints, which is commonly encountered in signal processing. 
	We have termed this generalized problem as the CMCQP. In general, the CMCQP is non-convex and difficult to solve, for which we have developed efficient solutions. To tackle the CMCQP, we have first reformulated it into an unconstrained optimization problem with respect to phase values of the studied variable only. 
	Then, we have identified two representative cases by means of studying the effect of the order and specific form of matrices embedded in the CMCQP. Based on this, we have devised a steepest descent/ascent method with fast determinations on its optimal step sizes, wherein a polynomial form of search function that leads to closed-form solutions of high accuracy has been derived. For completeness, we have also provided examples of recent applications associated with the two identified cases.
	Simulation results have verified the high accuracy of our proposed fast determinations on step sizes, and have shown the superiority of our proposed methods over existing ones in terms of different aspects.
	
	\appendix
	\subsection{Proof of \eqref{4gx3}} \label{Proofadd}
		\begin{proof}
			The explicit expressions of $\Re\{f(\boldsymbol{\theta})\}$, $\Im\{f(\boldsymbol{\theta})\}$, $\nabla\Re\{f(\boldsymbol{\theta})\}$, and $\nabla\Im\{f(\boldsymbol{\theta})\}$ in \eqref{4gori} are given as follows
			\begin{flalign}
				\Re\{f(&\boldsymbol{\theta})\}
				=\cos^{\mathrm{T}}(\boldsymbol{\theta})\Re\{\mathbf{A}\}\cos(\boldsymbol{\theta})-\cos^{\mathrm{T}}(\boldsymbol{\theta})\Im\{\mathbf{A}\}\sin(\boldsymbol{\theta})\nonumber\\
				&+\sin^{\mathrm{T}}(\boldsymbol{\theta})\Im\{\mathbf{A}\}\cos(\boldsymbol{\theta})+\sin^{\mathrm{T}}(\boldsymbol{\theta})\Re\{\mathbf{A}\}\sin(\boldsymbol{\theta})
				\label{eq:Real1}
				\\
				\Im\{f(&\boldsymbol{\theta})\}
				=\cos^{\mathrm{T}}(\boldsymbol{\theta})\Im\{\mathbf{A}\}\cos(\boldsymbol{\theta})+\cos^{\mathrm{T}}(\boldsymbol{\theta})\Re\{\mathbf{A}\}\sin(\boldsymbol{\theta})\nonumber\\
				&-\sin^{\mathrm{T}}(\boldsymbol{\theta})\Re\{\mathbf{A}\}\cos(\boldsymbol{\theta})+\sin^{\mathrm{T}}(\boldsymbol{\theta})\Im\{\mathbf{A}\}\sin(\boldsymbol{\theta})\label{eq:Imaginary1}
		   \end{flalign}   
	   
	   	   \begin{align}
	   	   	\nonumber\\[-11mm]
				\nabla\Re\{f(&\boldsymbol{\theta})\}=\Im\big\{\mathbf{A}-\mathbf{A}^{\mathrm{T}}\big\}\cos(\boldsymbol{\theta})\odot\cos{(\boldsymbol{\theta})}+\Re\big\{\mathbf{A} +\mathbf{A}^{\mathrm{T}}\big\}\nonumber\\
				&\times\sin(\boldsymbol{\theta})\odot\cos{(\boldsymbol{\theta})} -\Re\big\{\mathbf{A}+\mathbf{A}^{\mathrm{T}}\big\}\cos(\boldsymbol{\theta}) \odot\sin{(\boldsymbol{\theta})}\nonumber\\
				&+\Im\big\{\mathbf{A}-\mathbf{A}^{\mathrm{T}}\big\}\sin(\boldsymbol{\theta})\odot\sin{(\boldsymbol{\theta})}
				\label{subggA}\\
				\nabla\Im\{f(&\boldsymbol{\theta})\}=\Im\big\{\mathbf{A}+\mathbf{A}^{\mathrm{T}}\big\}\sin(\boldsymbol{\theta})\odot\cos{(\boldsymbol{\theta})}-\Re\big\{\mathbf{A} -\mathbf{A}^{\mathrm{T}}\big\}\nonumber\\
				&\times\cos(\boldsymbol{\theta})\odot\cos{(\boldsymbol{\theta})}-\Re\big\{\mathbf{A}-\mathbf{A}^{\mathrm{T}}\big\}\sin(\boldsymbol{\theta}) \odot\sin{(\boldsymbol{\theta})}\nonumber\\
				&-\Im\big\{\mathbf{A}+\mathbf{A}^{\mathrm{T}}\big\}\cos(\boldsymbol{\theta})\odot\sin{(\boldsymbol{\theta})}.
				\label{subggB}
			\end{align}
			Substituting \eqref{eq:Real1}, \eqref{eq:Imaginary1}, \eqref{subggA}, and \eqref{subggB} into \eqref{4gori}, using \eqref{relationship} and also the elementary properties  $\Re\big\{\mathbf{A}^{\mathrm{T}}\big\}=\Re\big\{\mathbf{A}^{\mathrm{H}}\big\}$ and $\Im\big\{\mathbf{A}^{\mathrm{T}}\big\}=-\Im\big\{\mathbf{A}^{\mathrm{H}}\big\}$, $\nabla\mathrm{Obj}_{\mathrm{I}}$ can then be written as \eqref{4gx3}. The proof is complete.
	\end{proof}
	
	\subsection{Proof of \eqref{4objt} with \eqref{mu3}--\eqref{mu1}} \label{ProofA}
	\begin{proof}
			The $3$rd-order Taylor expansions of the trigonometric functions in \eqref{Opt4} can be expressed as
			\begin{flalign}
				&\cos\big(\boldsymbol{\theta}^{(k)}+\tau\rho_{\mathrm{I}}\nabla^{(k)}\mathrm{Obj}_{\mathrm{I}}\big)
				=\tfrac{\tau\rho_{\mathrm{I}}^3}{6}\big|\nabla^{(k)}\mathrm{Obj}_{\mathrm{I}}\big|^3\odot\sin\big(\boldsymbol{\theta}^{(k)}\big)\nonumber\\
				&\;\;
				-\tfrac{\rho_{\mathrm{I}}^2}{2}\big|\nabla^{(k)}\mathrm{Obj}_{\mathrm{I}}\big|^2\odot\cos\big(\boldsymbol{\theta}^{(k)}\big)- \tau\nabla^{(k)}\mathrm{Obj}_{\mathrm{I}}\odot\sin\big(\boldsymbol{\theta}^{(k)}\big)\nonumber\\
				&\;\;
				\times\rho_{\mathrm{I}}+\cos\big(\boldsymbol{\theta}^{(k)}\big)
				+\mathcal{O}\big(\rho_{\mathrm{I}}^4\big)\label{cos}\\
				&
				\sin\big(\boldsymbol{\theta}^{(k)}+\tau\rho_{\mathrm{I}}\nabla^{(k)}\mathrm{Obj}_{\mathrm{I}}\big)=-\tfrac{\tau\rho_{\mathrm{I}}^3}{6}\big|\nabla^{(k)}\mathrm{Obj}_{\mathrm{I}}\big|^3\odot\cos\big(\boldsymbol{\theta}^{(k)}\big)
				\nonumber\\
				&\;\;
				-\tfrac{\rho_{\mathrm{I}}^2}{2}\big|\nabla^{(k)}\mathrm{Obj}_{\mathrm{I}}\big|^2\odot\sin\big(\boldsymbol{\theta}^{(k)}\big)+ \tau\nabla^{(k)}\mathrm{Obj}_{\mathrm{I}}\odot\cos\big(\boldsymbol{\theta}^{(k)}\big)
				\nonumber\\
				&\;\;
				\times\rho_{\mathrm{I}}+\sin\big(\boldsymbol{\theta}^{(k)}\big)+\mathcal{O}\big(\rho_{\mathrm{I}}^4\big).
				\label{sin}
			\end{flalign}
			
			Let $\widetilde{\boldsymbol{\Theta}}^{(k)}\triangleq\big[\cos^{\mathrm{T}}\!\big(\boldsymbol{\theta}^{(k)}\big),\, \sin^{\mathrm{T}}\!\big(\boldsymbol{\theta}^{(k)}\big)\big]^{\mathrm{T}}$ and
			\begin{align}
				\mathbf{A}_\mathrm{R}&\triangleq\begin{bmatrix}
					\Re\{\mathbf{A}\}&-\Im\{\mathbf{A}\}\\
					\Im\{\mathbf{A}\}&\Re\{\mathbf{A}\}
				\end{bmatrix}\\
				\mathbf{A}_\mathrm{I}&\triangleq\begin{bmatrix}
					\Im\{\mathbf{A}\}&\Re\{\mathbf{A}\}\\
					-\Re\{\mathbf{A}\}&\Im\{\mathbf{A}\}
				\end{bmatrix}.
			\end{align}	
			Substituting \eqref{cos} and \eqref{sin} into the objective of \eqref{Opt4}, we can firstly obtain
			\begin{align}
				\widetilde{\mathrm{Obj}}^{(k)}_{\mathrm{I}}(\rho_{\mathrm{I}})=\big|\tau\widetilde{\lambda}^{(k)}\rho_{\mathrm{I}}^3+\widetilde{\mu}^{(k)}\rho_{\mathrm{I}}^2+\tau\widetilde{\upsilon}^{(k)}\rho_{\mathrm{I}}+\widetilde{\eta}^{(k)} +\mathcal{O}\big(\rho_{\mathrm{I}}^4\big)\big|^2 \label{eta}
		\end{align}
		where the coefficients $\widetilde{\lambda}^{(k)}$, $\widetilde{\mu}^{(k)}$, $\widetilde{\upsilon}^{(k)}$, and $\widetilde{\eta}^{(k)}$ are respectively given by
		\begin{align}
			&\widetilde{\lambda}^{(k)}=\tfrac{1}{2}\big(\big|\mathbf{1}_2\otimes\nabla^{(k)}\mathrm{Obj}_{\mathrm{I}}\big|^2\odot\widetilde{\boldsymbol{\Theta}}^{(k)}\big)^{\mathrm{T}}\big(\mathbf{A}_{\mathrm{I}}-j\mathbf{A}_{\mathrm{R}}\big)\big(\mathbf{1}_2\nonumber\\
			&\ \; \otimes\nabla^{(k)}\mathrm{Obj}_{\mathrm{I}}\odot\widetilde{\boldsymbol{\Theta}}^{(k)}\big)-\tfrac{1}{2}\big(\mathbf{1}_2\otimes\nabla^{(k)}\mathrm{Obj}_{\mathrm{I}}\odot\widetilde{\boldsymbol{\Theta}}^{(k)}\big)^{\mathrm{T}}\nonumber\\
			&\ \; \times\big(\mathbf{A}_{\mathrm{I}}-j\mathbf{A}_{\mathrm{R}}\big)\big(\big|\mathbf{1}_2\otimes\nabla^{(k)}\mathrm{Obj}_{\mathrm{I}}\big|^2\odot\widetilde{\boldsymbol{\Theta}}^{(k)}\big)-\tfrac{1}{6}\big(\big|\mathbf{1}_2\nonumber\\
			&\ \; \otimes\nabla^{(k)}\mathrm{Obj}_{\mathrm{I}}\big|^3\odot\widetilde{\boldsymbol{\Theta}}^{(k)}\big)^{\mathrm{T}}\big(\mathbf{A}_{\mathrm{I}}-j\mathbf{A}_{\mathrm{R}}\big)\widetilde{\boldsymbol{\Theta}}^{(k)}+\tfrac{1}{6}\big(\widetilde{\boldsymbol{\Theta}}^{(k)}\big)^{\mathrm{T}}
			\nonumber\\
			&\ \; \times\big(\mathbf{A}_{\mathrm{I}}-j\mathbf{A}_{\mathrm{R}}\big)\big(\big|\mathbf{1}_2\otimes\nabla^{(k)}\mathrm{Obj}_{\mathrm{I}}\big|^3\odot\widetilde{\boldsymbol{\Theta}}^{(k)}\big)\label{etapr3}
		\end{align}
	
		\begin{align}
			\nonumber\\[-11mm]
			&\widetilde{\mu}^{(k)}=\big(\mathbf{1}_2\otimes\nabla^{(k)}\mathrm{Obj}_{\mathrm{I}}\odot\widetilde{\boldsymbol{\Theta}}^{(k)}\big)^{\mathrm{T}}\big(\mathbf{A}_{\mathrm{R}}+j\mathbf{A}_{\mathrm{I}}\big)\nonumber\\
			&\ \; \times\big(\mathbf{1}_2\otimes\nabla^{(k)}\mathrm{Obj}_{\mathrm{I}}\odot\widetilde{\boldsymbol{\Theta}}^{(k)}\big)-\tfrac{1}{2}\big(\big|\mathbf{1}_2\otimes\nabla^{(k)}\mathrm{Obj}_{\mathrm{I}}\big|^2\nonumber\\
			&\ \; \odot\widetilde{\boldsymbol{\Theta}}^{(k)}\big)^{\mathrm{T}}\big(\mathbf{A}_{\mathrm{R}}+j\mathbf{A}_{\mathrm{I}}\big)\widetilde{\boldsymbol{\Theta}}^{(k)}-\tfrac{1}{2}\big(\widetilde{\boldsymbol{\Theta}}^{(k)}\big)^{\mathrm{T}}\big(\mathbf{A}_{\mathrm{R}}+j\mathbf{A}_{\mathrm{I}}\big)\nonumber\\
			&\ \; \times\big(\big|\mathbf{1}_2\otimes\nabla^{(k)}\mathrm{Obj}_{\mathrm{I}}\big|^2\odot\widetilde{\boldsymbol{\Theta}}^{(k)}\big)\label{etapr2}\\
			&\widetilde{\upsilon}^{(k)}=\big(\mathbf{1}_2\otimes\nabla^{(k)}\mathrm{Obj}_{\mathrm{I}}\odot\widetilde{\boldsymbol{\Theta}}^{(k)}\big)^{\mathrm{T}}\big(\mathbf{A}_{\mathrm{I}}-j\mathbf{A}_{\mathrm{R}}\big)\widetilde{\boldsymbol{\Theta}}^{(k)}\nonumber\\
			&\ \;
			-\big(\widetilde{\boldsymbol{\Theta}}^{(k)}\big)^{\mathrm{T}}\big(\mathbf{A}_{\mathrm{I}}-j\mathbf{A}_{\mathrm{R}}\big)\big(\mathbf{1}_2\otimes\nabla^{(k)}\mathrm{Obj}_{\mathrm{I}}\odot\widetilde{\boldsymbol{\Theta}}^{(k)}\big)\label{etapr1}\\
			&\widetilde{\eta}^{(k)}=\big(\widetilde{\boldsymbol{\Theta}}^{(k)}\big)^{\mathrm{T}}\big(\mathbf{A}_{\mathrm{R}}+j\mathbf{A}_{\mathrm{I}}\big)\widetilde{\boldsymbol{\Theta}}^{(k)}.\label{etapr0}
		\end{align}

		Expanding the modulus and the square in \eqref{eta}, $\widetilde{\mathrm{Obj}}^{(k)}_{\mathrm{I}}(\rho_{\mathrm{I}})$ can be further derived as
	    \begin{align}
	    	\widetilde{\mathrm{Obj}}&^{(k)}_{\mathrm{I}}(\rho)
	    	=\Re\Big\{2\Big(\big(\widetilde{\eta}^{(k)}\big)^*\widetilde{\lambda}^{(k)}+\big(\widetilde{\mu}^{(k)}\big)^*\widetilde{\upsilon}^{(k)}\Big) \tau\rho^3\nonumber\\
	    	&+\Big(2\big(\widetilde{\eta}^{(k)}\big)^*\widetilde{\mu}^{(k)}+\big(\widetilde{\upsilon}^{(k)}\big)^*\widetilde{\upsilon}^{(k)}\Big) \rho^2+2\big(\widetilde{\eta}^{(k)}\big)^*\widetilde{\upsilon}^{(k)} \tau\rho\nonumber\\
	    	&+
	    	\big(\widetilde{\eta}^{(k)}\big)^*\widetilde{\eta}^{(k)}+\mathcal{O}\big(\rho^4\big)\Big\}.\label{4objtpr}
	    \end{align}
		Comparing \eqref{4objtpr} with \eqref{4objt}, we obtain the following results
		\begin{align}
			\lambda^{(k)}_{\mathrm{I}}&=2\Re\Big\{\big(\widetilde{\eta}^{(k)}\big)^*\widetilde{\lambda}^{(k)}+\big(\widetilde{\mu}^{(k)}\big)^*\widetilde{\upsilon}^{(k)}\Big\}\label{mupr3}\\
			\mu^{(k)}_{\mathrm{I}}&=\Re\Big\{2\big(\widetilde{\eta}^{(k)}\big)^*\widetilde{\mu}^{(k)}+\big(\widetilde{\upsilon}^{(k)}\big)^*\widetilde{\upsilon}^{(k)}\Big\}\label{mupr2}\\
			\upsilon^{(k)}_{\mathrm{I}}&=2\Re\Big\{\big(\widetilde{\eta}^{(k)}\big)^*\widetilde{\upsilon}^{(k)}\Big\}.\label{mupr1}
		\end{align}
		
		After substituting \eqref{etapr3}--\eqref{etapr0} into \eqref{mupr3}--\eqref{mupr1}, using \eqref{relationship} and \eqref{4gx3} together with some elementary properties of the Hadamard product, we can then simplify $\lambda^{(k)}_{\mathrm{I}}$, $\mu^{(k)}_{\mathrm{I}}$, and $\upsilon^{(k)}_{\mathrm{I}}$ as \eqref{mu3}, \eqref{mu2}, and \eqref{mu1}, respectively. The proof is complete.
	\end{proof}
	
	\subsection{Proof of \eqref{2objt} with \eqref{l23}--\eqref{l21}} \label{ProofB}
	\begin{proof}		
		Using \eqref{cos} and \eqref{sin} with $\nabla^{(k)}\mathrm{Obj}_{\mathrm{I}}$ replaced by $\nabla^{(k)}\mathrm{Obj}_{\mathrm{II}}$, the three coefficients in the third-order expansion of $\widetilde{\mathrm{Obj}}^{(k)}_{\mathrm{II}}(\rho_{\mathrm{II}})$ can be firstly written as
		\begin{align}
			\lambda^{(k)}_{\mathrm{II}}&=\big(\big|\mathbf{1}_2\otimes\nabla^{(k)}\mathrm{Obj}_{\mathrm{II}}\big|^2\odot\widetilde{\boldsymbol{\Theta}}^{(k)}\big)^{\mathrm{T}}\mathbf{A}_\mathrm{I}\big(\mathbf{1}_2\otimes\nabla^{(k)}\mathrm{Obj}_{\mathrm{II}}\nonumber\\
			&
			\odot\widetilde{\boldsymbol{\Theta}}^{(k)}\big)-\tfrac{1}{3}\big(\big|\mathbf{1}_2\otimes\nabla^{(k)}\mathrm{Obj}_{\mathrm{II}}\big|^3\odot\widetilde{\boldsymbol{\Theta}}^{(k)}\big)^{\mathrm{T}}\mathbf{A}_\mathrm{I}\widetilde{\boldsymbol{\Theta}}^{(k)}\label{lpr3}\!\!\!\\
			\mu^{(k)}_{\mathrm{II}}&=\big(\mathbf{1}_2\otimes\nabla^{(k)}\mathrm{Obj}_{\mathrm{II}}\odot\widetilde{\boldsymbol{\Theta}}^{(k)}\big)^{\mathrm{T}}\mathbf{A}_\mathrm{R}\big(\mathbf{1}_2\otimes\nabla^{(k)}\mathrm{Obj}_{\mathrm{II}}\nonumber\\
			&
			\odot\widetilde{\boldsymbol{\Theta}}^{(k)}\big)-\big(\big|\mathbf{1}_2\otimes\nabla^{(k)}\mathrm{Obj}_{\mathrm{II}}\big|^2\odot\widetilde{\boldsymbol{\Theta}}^{(k)}\big)^{\mathrm{T}}\mathbf{A}_\mathrm{R}\widetilde{\boldsymbol{\Theta}}^{(k)}\label{lpr2}\!\\
			\upsilon^{(k)}_{\mathrm{II}}&=2\big(\mathbf{1}_2\otimes\nabla^{(k)}\mathrm{Obj}_{\mathrm{II}}\odot\widetilde{\boldsymbol{\Theta}}^{(k)}\big)^{\mathrm{T}}\mathbf{A}_\mathrm{I}\widetilde{\boldsymbol{\Theta}}^{(k)}\label{lpr1}
		\end{align}
		where the Hermitian property of $\mathbf{A}$ has been used. Applying \eqref{relationship} and the elementary operations on real and imaginary values, \eqref{lpr3}, \eqref{lpr2}, and \eqref{lpr1} can be respectively rewritten as
		\begin{align}
			\lambda^{(k)}_{\mathrm{II}}=\;&\Im\Big\{\big(\big|\nabla^{(k)}\mathrm{Obj}_{\mathrm{II}}\big|^2\odot\mathbf{x}^{(k)}\big)^{\mathrm{H}}\mathbf{A}\big(\nabla^{(k)}\mathrm{Obj}_{\mathrm{II}}\odot\mathbf{x}^{(k)}\big)\nonumber\\
			&-\tfrac{1}{3}\big(\big|\nabla^{(k)}\mathrm{Obj}_{\mathrm{II}}\odot\mathbf{x}^{(k)}\big|^2\big)^{\mathrm{H}}\mathbf{A}\mathbf{x}^{(k)}\Big\}\label{2lpr3}\\
			\mu^{(k)}_{\mathrm{II}}=\;&\Re\Big\{\big(\nabla^{(k)}\mathrm{Obj}_{\mathrm{II}}\odot\mathbf{x}^{(k)}\big)^{\mathrm{H}}\mathbf{A}\big(\nabla^{(k)}\mathrm{Obj}_{\mathrm{II}}\odot\mathbf{x}^{(k)}\big)\nonumber\\
			&-\big(\big|\nabla^{(k)}\mathrm{Obj}_{\mathrm{II}}\big|^2\odot\mathbf{x}^{(k)}\big)^{\mathrm{H}}\mathbf{A}\mathbf{x}^{(k)}\Big\}\label{2lpr2}\\
			\upsilon^{(k)}_{\mathrm{II}}=\;&2\Im\Big\{\big(\nabla^{(k)}\mathrm{Obj}_{\mathrm{II}}\odot\mathbf{x}^{(k)}\big)^{\mathrm{H}}\mathbf{A}\mathbf{x}^{(k)}\Big\}.\label{2lpr1}
		\end{align}
		
		Using \eqref{g2} and some elementary properties of the Hadamard product, \eqref{2lpr3}, \eqref{2lpr2} and \eqref{2lpr1} can be simplified as \eqref{l23}, \eqref{l22}, and \eqref{l21}, respectively. The proof is complete.
	\end{proof}
	


\end{document}